\documentclass[aps,pra,amsmath,amssymb, twocolumn, showpacs]{revtex4-1}

\usepackage{graphicx} 
\usepackage{dcolumn}
\usepackage{bm}
\usepackage{amsmath}
\usepackage{amssymb}
\usepackage{color}
\usepackage{bbold}
\usepackage{bm}
\usepackage{mwe}
\usepackage{diagbox}
\usepackage{physics}

\newcommand{\cl}[1]{\mathcal{#1}}

\newcommand{\ave}[1]{\expval{#1}}

\newcommand{\llangle}[1][]{\savebox{\@brx}{\(\m@th{#1\langle}\)}
  \mathopen{\copy\@brx\kern-0.5\wd\@brx\usebox{\@brx}}}
\newcommand{\rrangle}[1][]{\savebox{\@brx}{\(\m@th{#1\rangle}\)}
  \mathclose{\copy\@brx\kern-0.5\wd\@brx\usebox{\@brx}}}

\newcommand{\up}{\uparrow}
\newcommand{\dn}{\downarrow}

\usepackage{textcomp}

\newcommand{\um}{\textmu m}

\newcommand{\caf}{$^{40}\text{Ca}^{+}$}
\newcommand{\cafthree}{$^{43}\text{Ca}^{+}$}

\begin{document}

\title{Engineering generalized Gibbs ensembles with trapped ions}

\author{Florentin Reiter$^{1,2}$}
\thanks{These two authors contributed equally}
\author{Florian Lange$^{3}$}
\thanks{These two authors contributed equally}
\author{Shreyans Jain$^{2}$}
\author{Matt Grau$^{2}$}
\author{Jonathan P. Home$^{2}$}
\author{Zala Lenar\v{c}i\v{c}$^4$}

\affiliation{$^1$Harvard University, Department of Physics, Cambridge, MA 02138, USA}
\affiliation{$^2$Institute for Quantum Electronics, ETH Z\"urich, Otto-Stern-Weg 1, 8093 Z\"urich, Switzerland}
\affiliation{$^3$Institute for Theoretical Physics, University of Cologne, D-50937 Cologne, Germany}
\affiliation{$^4$Department of Physics, University of California, Berkeley, California 94720, USA}

\begin{abstract}
The concept of generalized Gibbs ensembles (GGEs) has been introduced to describe steady states of integrable models. Recent advances show that GGEs can also be stabilized in nearly integrable quantum systems when driven by external fields and open.
Here, we present a weakly dissipative dynamics that drives towards a steady-state GGE and is realistic to implement in systems of trapped ions.
We outline the engineering of the desired dissipation by a combination of couplings which can be realized with ion-trap setups and discuss the experimental observables needed to detect a deviation from a thermal state. We present a novel mixed-species motional mode engineering technique in an array of micro-traps and demonstrate the possibility to use sympathetic cooling to construct many-body dissipators. Our work provides a blueprint for experimental observation of GGEs in open systems and opens a new avenue for quantum simulation of driven-dissipative quantum many-body problems.
\end{abstract}

\maketitle

\section{Introduction}
Providing a compact description of complicated many-body systems is a challenging task. Studies of equilibration of interacting quantum many-body systems have revealed that one can borrow statistical descriptions, taking into account conservation laws of equilibrating systems, to achieve that \cite{polkovnikov11, gogolin16}. If we suddenly excite an ergodic system, for which energy is the only conservation law, steady state expectation values will be given by a {\it Gibbs ensemble}
\begin{equation}
\rho_{\mathrm{th}}=\frac{e^{-\beta H_0}}{\tr[e^{-\beta H_0}]}
\end{equation}
with temperature $1/\beta$ determined by the amount of energy that has been injected into the system with the excitation process \cite{polkovnikov11}.
Such a description by the single parameter -- temperature -- is, therefore, an incredible simplification for an  interacting model with a priori exponentially many degrees of freedom.

\begin{figure}[t]
\includegraphics[width=0.475\textwidth]{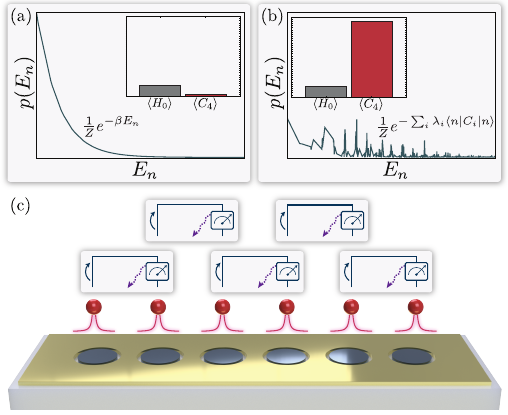}
\caption{
\label{fig::sketch}
Activating integrability.
(a) Following an excitation of a non-integrable or nearly integrable system, the steady state exhibits thermal behavior with small expectation values of operators other than the Hamiltonian.
(b) A nearly integrable system with a weak dissipative drive will show highly non-thermal behavior with a distribution $p(E_n)$ approximately described by a generalized Gibbs ensemble (GGE) and large expectation values for conservation laws, e.g., $\expval{C_4}$.
(c) A one dimensional array of trapped-ions has interactions which decay as a power-law, and is thus nearly integrable for large exponents. We re-activate integrable effects by engineering the dissipation, which stabilizes a GGE as the steady state.
}
\end{figure}

Similarly, {\it generalized Gibbs ensembles} (GGE) were proposed to describe local observables in steady states of systems with additional local conservation laws $C_i$.
The existence of additional conservation laws highly restricts the dynamics and prevents the system from thermalizing. 
GGEs have a form related to that of a Gibbs ensemble \cite{rigol07},
\begin{equation}\label{EqGGE}
\rho_{\textrm{GGE}} = \frac{e^{-\sum_i\lambda_i C_i}}{ \tr[e^{- \sum_i \lambda_i C_i}]}.
\end{equation}
but with additional Lagrange multipliers $\lambda_i$ associated with additional conserved quantities $C_i$ that all commute with the Hamiltonian, $[C_i,H_0]=0$. 
Exemplary models with macroscopically many local conservation laws, where applicability of GGE has been widely studied theoretically, are Bethe-ansatz-solvable and non-interacting integrable systems
\cite{essler16, vidmar16, cazalilla16, caux16, calabrese07gge, cramer08gge, barthel08gge, essler12gge, pozsgay13, fagotti13gge2, fagotti13, wouters14, pozsgay14a, sotiriadis14gge, goldstein14gge, brockmann14gge, rigol14gge, mestyan15gge, ilievski15a, eisert15gge, ilievski16gge, piroli16gge}.
In this case, Lagrange parameters $\lambda_i$ are fixed by the knowledge of the initial state $\ket{\psi(0)}$, as well as $\ave{C_i}=const$, 
\begin{equation}\label{EqClosed}
\ave{\psi(0)| C_k | \psi(0)}\overset{!}{=}
\tr\left[C_k \frac{e^{-\sum_i\lambda_i C_i}}{\tr[e^{-\sum_i\lambda_i C_i}]}\right], \quad \forall k.	
\end{equation}
Since generically $\lambda_i\neq0$, integrable systems remain non-thermal up to arbitrary times. 

Applicability of GGEs was confirmed also experimentally in a cold atoms setup~\cite{langen15} where, up to some time, a closed and integrable system can be prepared. Ref.~\cite{langen15} showed that GGEs for a Lieb-Liniger model can provide an accurate description of an interacting trapped 1D Bose gas. 
However, it is very difficult to simulate integrable systems due to their fine-tuned nature: adding practically any other terms to an integrable model will break integrability and cause eventual thermalization \cite{bertini15,bertini16,mallayya18a,mallayya18,durnin20,Tang18, li18,caux19,schemmer19,moller20}. 
Therefore it was believed that in realistic systems traces of integrability can be seen only in the transient dynamics \cite{moeckel08,kollar11,Barnett2011,gring12,Mitra2013,Marcuzzi2013,essler14,NessiIucci2014,Demler2015,canovi16} while the steady state is always thermal due to realistic integrability breaking terms.

In recent works \cite{lange17,lange18,lenarcic18}, Lange and Lenar\v{c}i\v{c} demonstrated that properties related to integrability are not as fragile as previously believed: if one weakly drives an only approximately integrable system and at the same time allows it to cool via a weak coupling to the environment,  the system will nonetheless relax to a steady state approximated with a generalized Gibbs ensemble, supplemented with a small correction $\delta\rho$,
\begin{equation}\label{EqRhoNESS-0}
\rho_{\infty}\equiv\lim_{t\to\infty}\rho(t) = \rho_{\textrm{GGE}} + \delta\rho.
\end{equation}
The major difference to the closed strictly integrable setup, Eq.~\eqref{EqClosed}, is that here the Lagrange multipliers $\lambda_i$ are determined by the integrability breaking perturbations themselves, through a stationarity condition \cite{lange17, lange18, lenarcic18}
\begin{equation}\label{eq::lambdaCondOpen}
\partial_t \ave{C_k} \approx \tr\left[C_k \cl{L}_p \frac{e^{-\sum_i\lambda_i C_i}}{\tr[e^{-\sum_i\lambda_i C_i}]}\right] \overset{!}{=}0, \quad \forall k.
\end{equation}
Here $\cl{L}_p$ denotes the Liouville operator corresponding to perturbations which weakly break the integrability, and as a consequence the conservation laws, while driving and cooling the system. 
Such a setup is much more versatile because it does not require the fine-tuned perfect integrability and at the same time allows for the engineering of GGEs through a particular choice of perturbations.
A remarkable consequence is that one can stabilize steady states with large expectation values of nearly conserved operators 
$\ave{C_{i}}$, if a large corresponding $\lambda_i$ is established. For example, in solid-state spin chain materials, approximately described by an XXZ model coupled to phonons, a weak laser driving could stabilize steady states with huge heat and spin currents, since these are (partial) conservation laws of the XXZ model \cite{lange17}. 
Alternatively, driving and openness can be provided by Markovian dissipative processes \cite{lange18}. As shown in Fig.~\ref{fig::sketch}, the latter could be experimentally realized with trapped-ion platforms where long-range interaction between ions inevitably breaks integrability. By adding engineered dissipative processes, a GGE ensemble would be stabilized.
As a consequence, the distribution 
$p(E_n) = {\exp({-\sum_i \lambda_i \ave{n|C_i|n}})}/{\tr[e^{-\sum_i \lambda_i C_i}]}$ over the eigenstates of $H_0\ket{n}=E_n\ket{n}$
 would differ from a thermal one $p(E_n) = {\exp({-\beta E_n})}/{\tr[e^{-\beta H_0}]}$. While $p(E_n)$ is essentially impossible to measure experimentally, the non-thermal nature of $\rho_{\textrm{GGE}}$ is more easily detected through possible exceedingly large expectation values of conserved operators $\ave{C_i}$ of parent XY or transverse-Ising Hamiltonian.

In this work, we address the latter option and discuss an implementation based on ion-trap technology \cite{Barreiro2011, Lin2013, Leibfried2003}.
Controllable coherent couplings in ion-trap systems have been widely used for the realization of spin models \cite{Porras2004, Deng2005}, and numerous milestone experiments have been conducted on a variety of systems \cite{Friedenauer2008, Kim2010, Britton2012, Islam2013, Jurcevic2014, Richerme2014, Bohnet2016, Zhang2017, Kokail2019}.
State-of-the-art Paul traps \cite{Lin2013, Jurcevic2014, Zhang2017}, Penning traps \cite{Bohnet2016, Safavi2018, Jordan2019}, and micro-traps \cite{Cirac2000, Wilson2014, Yang2016, Mielenz2016, Jain2018} offer a rich toolbox of couplings suitable to engineer coherent Hamiltonian, as well as dissipative interactions. Sympathetic cooling based on mixed-species ion chains is well-studied in Paul traps \cite{Barrett2003, Guggemos2015, Negnevitsky2018} where it is used to remove entropy from the motional modes of the ions.
Here, we realize a driven-dissipative dynamics consisting of a spin Hamiltonian, in combination with one- and two-body dissipation. The dissipators are engineered combining tunable carrier and sideband couplings with repumper drives or sympathetic cooling as sources of dissipation. To tightly confine the motional modes to the interacting particles, we propose a novel mixed-species mode engineering technique that can be realized in micro-trap arrays. The resulting dynamics stabilizes a steady state approximately described by a generalized Gibbs ensemble, despite different integrability breaking terms. 

The paper is organized the following way: In Sec.~\ref{sec::Model} we introduce one choice of a Hamiltonian and Lindblad operators that could be realized in a trapped ion experiment. In Sec.~\ref{sec::NESS} we present numerical results and discuss experimental signatures and means to measure that a GGE approximates the stabilized steady state. In Section \ref{SecImp}--\ref{sec:generalization} we present the engineering of the elementary dissipators. We then scale up the interactions in Sec.~\ref{sec:scalability} and Sec.~\ref{sec:micro}, where we discuss the mode engineering in arrays of micro-traps.

\section{Model}\label{sec::Model}
The theory of activating integrability and engineering steady states described by generalized Gibbs ensembles in realistic systems is generic and applies to different systems approximately described by an integrable model, e.g., transverse field Ising or XXZ Heisenberg chain, Lieb-Liniger or Tonks-Girardeau Bose 1D gas. We choose an integrable model that is closest to the state-of-the art trapped ions setups. We consider the XY-Hamiltonian in the presence of a magnetic field $h$,
\begin{align}\label{EqXYh}
H_\mathrm{XY}&=\sum_j  J_x \sigma_j^x \sigma_{j+1}^x + J_y \sigma_j^y \sigma_{j+1}^y + h \sigma_j^z,
\end{align}
which belongs to the class of non-interacting integrable models. Such a Hamiltonian can be implemented by using standard techniques developed in the trapped-ion field.
In contrast to Ref. \cite{Jurcevic2014}, we rotate the spin axes by $\pi/2$ around the y-axis. The resulting YZ-model will allow us to facilitate an experimental implementation of Lindblad terms. We consider
\begin{align}\label{EqYZh}
H_0&=\sum_j   J_z \sigma_j^z \sigma_{j+1}^z + J_y \sigma_j^y \sigma_{j+1}^y + h \sigma_j^x.
\end{align}
An alternative realization based on the XY-Hamiltonian in combination with sympathetic cooling is presented in Sec.~\ref{sec:generalization}.
 
In traditional setups with trapped ions, the coupling between spins that are $d$ sites apart actually decays as $d^{-\alpha} \sigma_j^{z(y)} \sigma_{j+d}^{z(y)}$ with $\alpha \in [2 ,3]$. This is one inevitable source of integrability breaking since such Hamiltonian can no longer be diagonalized via a Jordan-Wigner transformation, neither is it Bethe-Ansatz solvable. However, if the decay is fast enough one can consider such a system as nearly integrable. In our analysis we will take into consideration only the leading contribution
\begin{align}\label{EqH1}
H_1&=\epsilon_1\sum_j  J_z \sigma_j^z \sigma_{j+2}^z + J_y \sigma_j^y \sigma_{j+2}^y, \quad \epsilon_1=\frac{1}{2^\alpha}.
\end{align}
$H_1$ alone would thermalize the system, however, non-thermal steady states approximated by GGE can be achieved when a weak coupling to Lindblad non-equilibrium baths is added. Here, we will consider the homogeneous bulk dissipators of two types, $a=1,2$,
\begin{equation}
\cl{D}^{(a)}\rho=\sum_j L_j^{(a)} \rho {L_j^{(a)}}^\dagger - \frac{1}{2} \{ {L_j^{(a)}}^\dagger L_j^{(a)}, \rho \}
\end{equation}
with Lindblad operators at site $j$
\begin{align}
\label{EqLindZ1}
&L_j^{(1)}=\sqrt{\epsilon(1-\gamma)}\, S_{j}^{-}, \quad \gamma \in [0,1]  
\\
\label{EqLindZ2}
&L_j^{(2)}=\sqrt{\epsilon\gamma}\, S_{j}^+ P_{j+1}^{\downarrow},
\end{align}
Here $P_{j}^{\downarrow}=\frac{1}{2}(\mathbb{1}_j - \sigma^z_j)$ is a projection on the spin-down state at site $j$. 

The steady state density matrix $\rho_\infty$ is determined by
\begin{equation}\label{eq::exact}
\dot{\rho}_\infty=(\cl{L}_0 + \cl{L}_p)\rho_\infty=0,
\end{equation}
where $\cl{L}_0$ is a dominant term in the Liouvillian, while  
$\cl{L}_p=\cl{L}_u + \cl{L}_{m}$ captures unitary and Markovian perturbations,
\begin{align}\label{EqLiouv}
\cl{L}_0 \rho &= -i[H_0,\rho], \notag  \\ 
\cl{L}_u \rho &= -i\left[H_1,\rho\right],\notag  \\
\cl{\mathcal{L}}_m\rho&=  \left(  \cl{D}^{(1)} + \cl{D}^{(2)} \right) \rho. 
\end{align}
Despite the fact that the underlying model $H_0$ in Eq.~\eqref{EqXYh} is non-interacting, the next-nearest neighbor interaction $H_1$ and our choice of Lindblad operators hinders analytical solvability and requires a numerical
solution.
Since our Lindblad operators have local nature, we also cannot use recently proposed hydrodynamic description \cite{bastianello20}.

For a review of the theory of weakly driven nearly integrable systems, developed in \cite{lange17,lange18,lenarcic18} see App.~\ref{sec::Ensembles}. In the next section we go on to establish that this model does stabilize a steady state approximately described by a GGE ensemble, despite different sources of integrability breaking. 

\section{Experimental signatures}\label{sec::NESS}

The experiment we are proposing would aim to show that a highly non-thermal steady state, described with a generalized Gibbs ensemble, can be stabilized in a nearly integrable model given by $H_0+H_1$, Eqs.~(\ref{EqYZh},\ref{EqH1}), if weakly driven ($\epsilon\ll 1$) with dissipation Eqs.~(\ref{EqLindZ1},\ref{EqLindZ2}). 

\begin{figure}[t!]
\includegraphics[width=.9\linewidth]{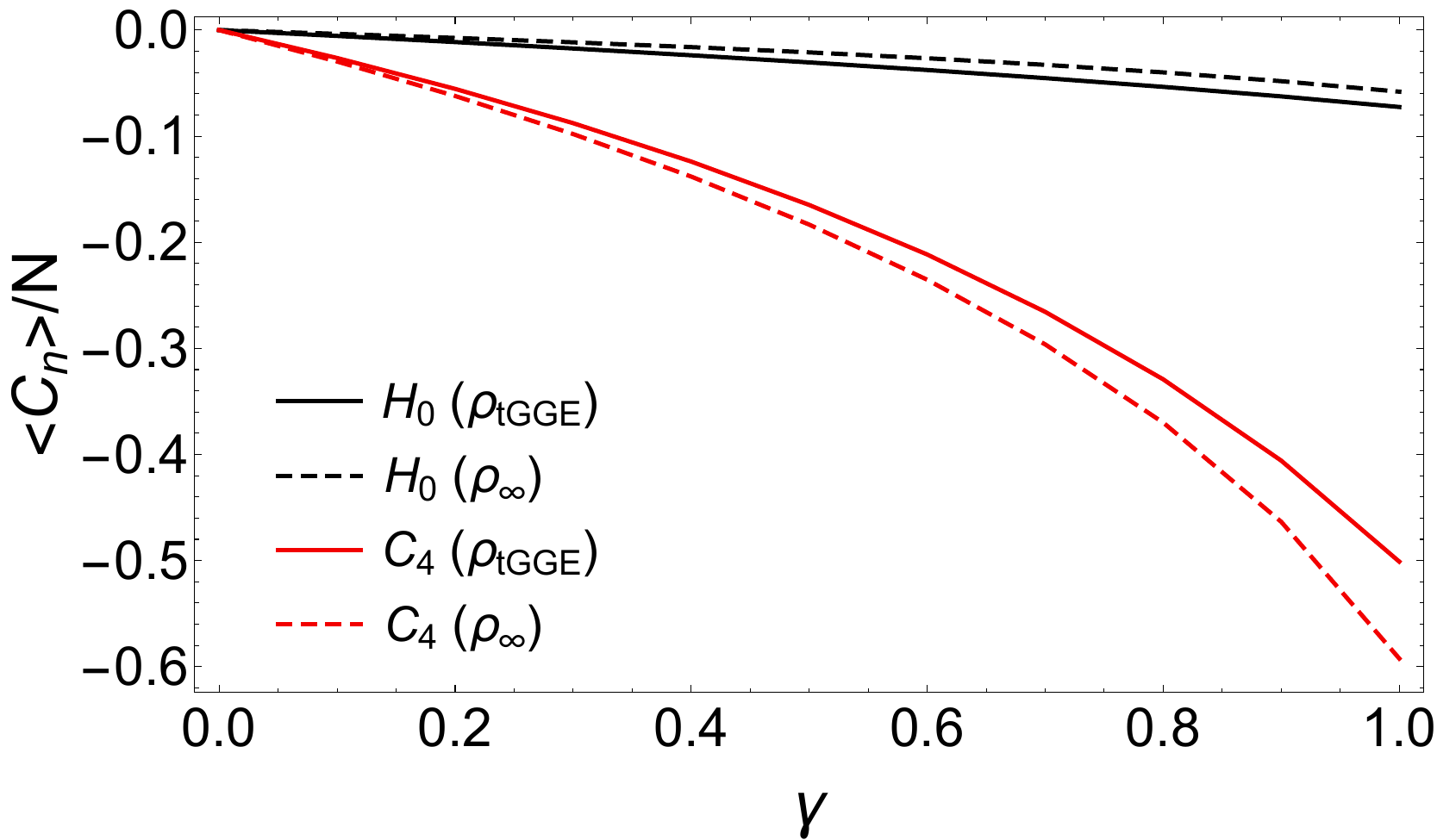}
\caption{$\ave{H_0}$ and $\ave{C_4}$ as a function of relative driving strength $\gamma$, Eqs.~(\ref{EqLindZ1},\ref{EqLindZ2}). A qualitative agreement between the exact $\rho_\infty$ and $\rho_\textrm{tGGE}$ confirms that a generalized Gibbs ensemble is stabilized. A large $\ave{C_4}$, which could not be obtained from a nearly infinite temperature steady state consistent with a small $\ave{H_0}$, indicates that a highly non-thermal GGE is stabilized. Parameters: $J_y=h=1$, $J_z=0.1$; $N=10, N_c=4$ for $\rho_\textrm{tGGE}$ and $N=6, \epsilon=0.01$ for $\rho_\infty$.
\label{FigExpVal}}
\end{figure}

To show that a GGE is stabilized by driving, we compare in Fig.~\ref{FigExpVal} the steady state expectation values obtained with the exact density matrix, Eq.~\eqref{eq::exact} or with a GGE, Eq.~\eqref{eq::tGGEMain}. A qualitative agreement confirms that a GGE is stabilized.
The first notable conclusion from the numerical analysis is that in the limit of small driving, $\epsilon\ll 1$, a truncated generalized Gibbs ensemble 
\begin{equation}\label{eq::tGGEMain}
\rho_{\textrm{tGGE}} = \frac{e^{-\sum_{i=1}^{N_c}\lambda_i C_i}}{ \tr[e^{- \sum_{i=1}^{N_c} \lambda_i C_i}]}
\end{equation}
parametrized with only $N_c=4$ Lagrange parameters can capture expectation values of local observables we are interested in, while other more complicated conservation laws can be neglected. In comparison, a full density matrix requires $4^N$ parameters in the system size $N$. This shows that a description in terms of truncated GGEs is extremely compact. For a more thorough comparison see App.~\ref{sec::Ensembles}.

Another interesting observation is that the next-nearest neighbors coupling $H_1$, Eq.~\eqref{EqH1}, does not have a strong impact on the steady state. While in a closed setup $H_1$ is crucial as it dictates relaxation towards a thermal state, in our setup it is dominated by Lindblad terms, see App.~\ref{sec::Ensembles} for details.  Therefore, we neglect it in results presented in the main text. 

A physically most interesting consequence of stabilizing a steady state approximately described with a GGE is that the expectation values of conservation laws, 
\begin{equation}\label{eq::evaCi}
\ave{C_k}=\tr\left[C_k \frac{e^{-\sum_i \lambda_i C_i}}{\tr[e^{-\sum_i \lambda_i C_i}]}\right],	
\end{equation}
can be much larger than in a thermal state,
$\tr[C_k e^{-\beta_0}/\tr[e^{-\beta H_0}]]$. 
In this respect conservation laws are measurably distinct from other operators: a generic observable $O$ will have a much smaller expectation value than a conservation law, $\tr[O\rho_\mathrm{GGE}]\ll \tr[C_k\rho_\mathrm{GGE}]$, given they have the same norm $\tr[O^2]=\tr[C_k^2]$.
The conservation law $C_k$ will show a particularly large expectation value if driving is such that it stabilizes a GGE with a large corresponding Lagrange parameter $\lambda_k$.
Which $\lambda_k\neq 0$ depends on the symmetry of the driving, i.e., the Lindblad operators \cite{lange17,lange18,yamamoto20}.
A strong response of conservation laws to weak driving could have practical implications, such as heat and spin pumping \cite{lange17} in spin chain materials, but can also serve to detect that a GGE has been stabilized despite the integrability breaking terms. In the following we discuss  ways to detect large expectation values of conserved (or partially conserved) operators, which could not be possible in a thermal state.

One possibility is to compare expectation values of observables which {\it do} or {\it do not} overlap with conservation laws. 
If driving stabilizes $\rho_\mathrm{GGE}$ with a large Langrange multiplier $\lambda_i$ associated with the conservation law $C_i$, observable $O$ that has a nonzero overlap with $C_i$, $\tr[OC_i]\neq 0$, will show a large expectation value.

\begin{figure}[t!]
\center
\includegraphics[width=.9\linewidth]{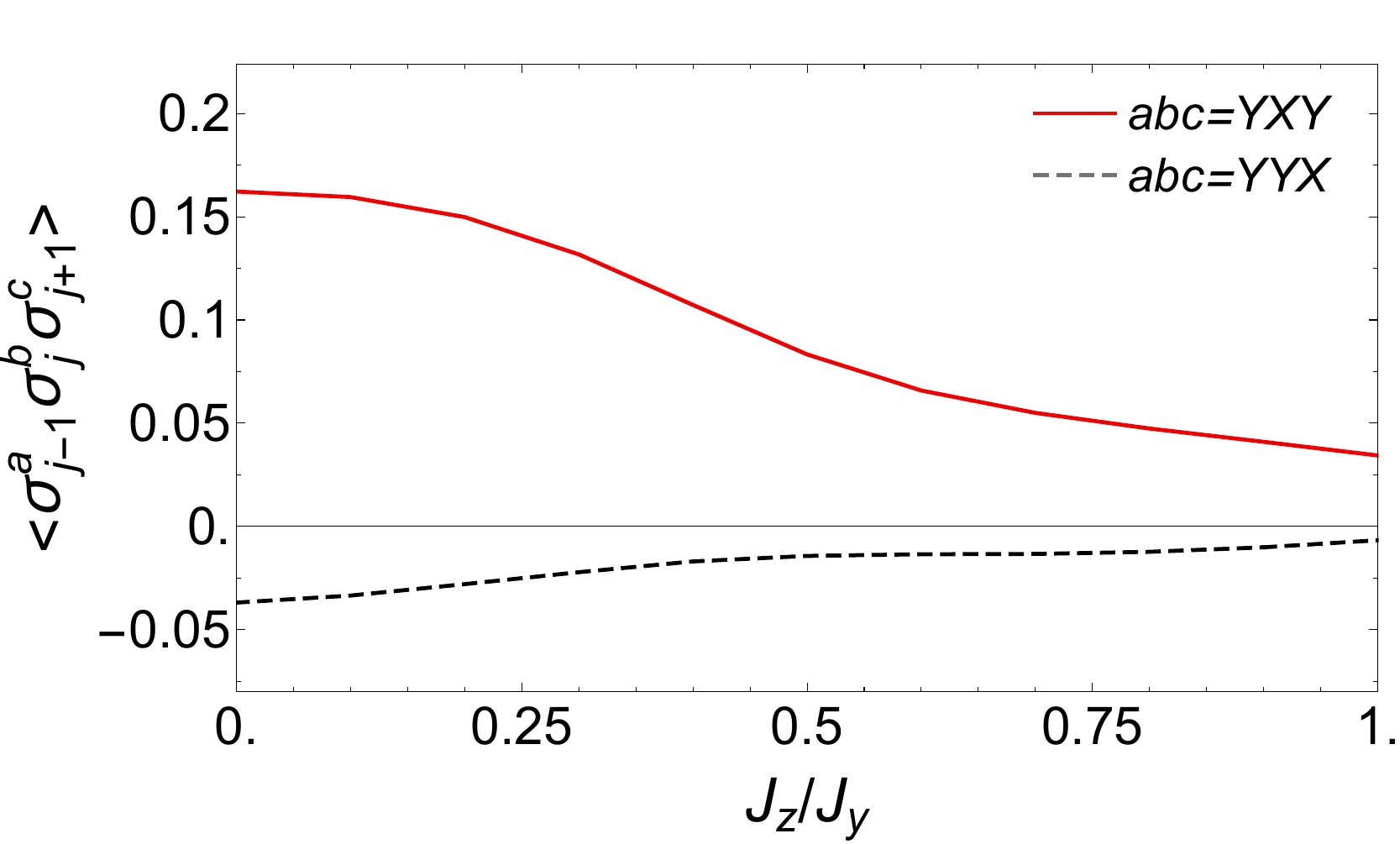}
\caption{Expectation values $\ave{\sigma^y_{j-1}\sigma^y_{j}\sigma^x_{j+1}}$ and $\ave{\sigma^y_{j-1}\sigma^x_{j}\sigma^y_{j+1}}$ as a function of anisotropy $J_z/J_y$ obtained from the exact steady state $\rho_\infty$ at $N=6, \epsilon=0.01$, $J_y=1$, $\gamma=0.8$. $\ave{\sigma^y_{j-1}\sigma^x_{j}\sigma^y_{j+1}}$ shows a much larger value because it is part of a conservation law, $C_4$, Eq.~\eqref{EqC4}.
\label{FigYYZ} }
\end{figure}

Let us, for example, consider $\ave{\sigma^y_{j-1}\sigma^x_{j}\sigma^y_{j+1}}\equiv\ave{\textrm{YXY}}$ and $\ave{\sigma^y_{j-1}\sigma^y_{j}\sigma^x_{j+1}}\equiv\ave{\textrm{YYX}}$. At least at small $J_z/J_y \ll 1$, where $H_0\approx J_y \sigma_j^y \sigma_{j+1}^y + h \sigma_j^x$, it is easy to estimate thermal expectation values $\ave{\textrm{XYX}}_{\textrm{th}}, \ave{\textrm{XXY}}_{\textrm{th}}$ via  expansion
$\rho_\mathrm{th}= e^{-\beta H_0}/Z \approx (\mathbb{1} - \beta H_0 + \beta^2 H_0^2/2 + \dots)/Z$:  a nonzero $\ave{\textrm{YYX}}_{\textrm{th}}$ is dominantly coming from the 2nd order in $\beta$, while $\ave{\textrm{YXY}}_{\textrm{th}}$  from the 3rd order. In a thermal state one would therefore expect $\ave{\textrm{YXY}}_{\textrm{th}} \ll \ave{\textrm{YYX}}_{\textrm{th}}$. However, our numerical result in Fig.~\ref{FigYYZ} shows $\ave{\textrm{YXY}} \gg \ave{\textrm{YYX}}$. This observation is a clear sign that the steady state is not thermal. The large expectation value of $\ave{\textrm{YXY}}$ is a direct consequence of the fact that this operator is part of $C_4$, Eq.\eqref{EqC4}, therefore its expectation value has also a linear contribution in $\lambda_4$ in the expansion, since driving stabilizes a GGE with $\lambda_4\neq0$.

In our setup, besides the Hamiltonian $H_0$ ($H_0=C_2$ according to the notation we use, see App.~\ref{sec::Ensembles}), the next simplest local extensive conservation law $C_i=\sum_j c^{(i)}_j$ that shows a strong response to the dissipative driving is
\begin{align}\label{EqC4}
C_4=\sum_j\sum_{\mu=z,y}
&J_\mu \sigma^\mu_j \sigma^x_{j+1} \sigma^x_{j+2} \sigma^\mu_{j+3}
-h \sigma^\mu_j \sigma^x_{j+1} \sigma^\mu_{j+2}\notag \\
& \hspace{3cm}+ J_{\bar{\mu}} \sigma^\mu_j \sigma^\mu_{j+1},
\end{align}
where $\bar{z}=y$ and $\bar{y}=z$. Fig.~\ref{FigExpVal} shows the comparison of $\ave{H_0}=\ave{C_2}$ and $\ave{C_4}$ as a function of the relative strength $\gamma$, Eqs.~(\ref{EqLindZ1},\ref{EqLindZ2}), obtained from a GGE calculation on $N=10$ sites or a full steady state density matrix on $N=6$ sites. While $\ave{H_0}(\gamma)$ shows rather small values, $\ave{C_4}(\gamma)$ is bigger. That would never be the case, if $\ave{C_4}(\gamma)$ was evaluated with respect to (an almost infinite temperature) thermal ensemble, which reproduces small $\ave{H_0}(\gamma)$. This observation alone suggests that a non-thermal steady state is stabilized.

In order to quantify how non-thermal the steady state is and to select optimal parameters, we introduce the ratio $\eta_O$ 
\begin{equation}\label{EqEta}
\eta_O=\frac{\tr[O \rho_x] - \tr[O \rho_\mathrm{th}]}{\tr[O \rho_\mathrm{th}]},
\end{equation}
calculated with respect to the exact steady state, $\rho_x= \rho_\infty$, or with the truncated GGE, $\rho_x= \rho_{\textrm{tGGE}}$. For calculations with $\rho_x= \rho_\infty$ we define $\rho_\mathrm{th}$ as a thermal state with respect to $H_0$, Eq.~\eqref{EqXYh}, with temperature determined from the condition 
$\tr[H_0 \rho_\infty]=\tr[H_0 e^{-\beta H_0}/\tr[e^{-\beta H_0}]]$.
For calculations based on $\rho_x= \rho_{\textrm{tGGE}}$ the temperature in $\rho_\mathrm{th}$ is calculated using a Gibbs ensemble ansatz with $H_0$ as the only conservation law.

\begin{figure}[t!]
\center
\includegraphics[width=.9\linewidth]{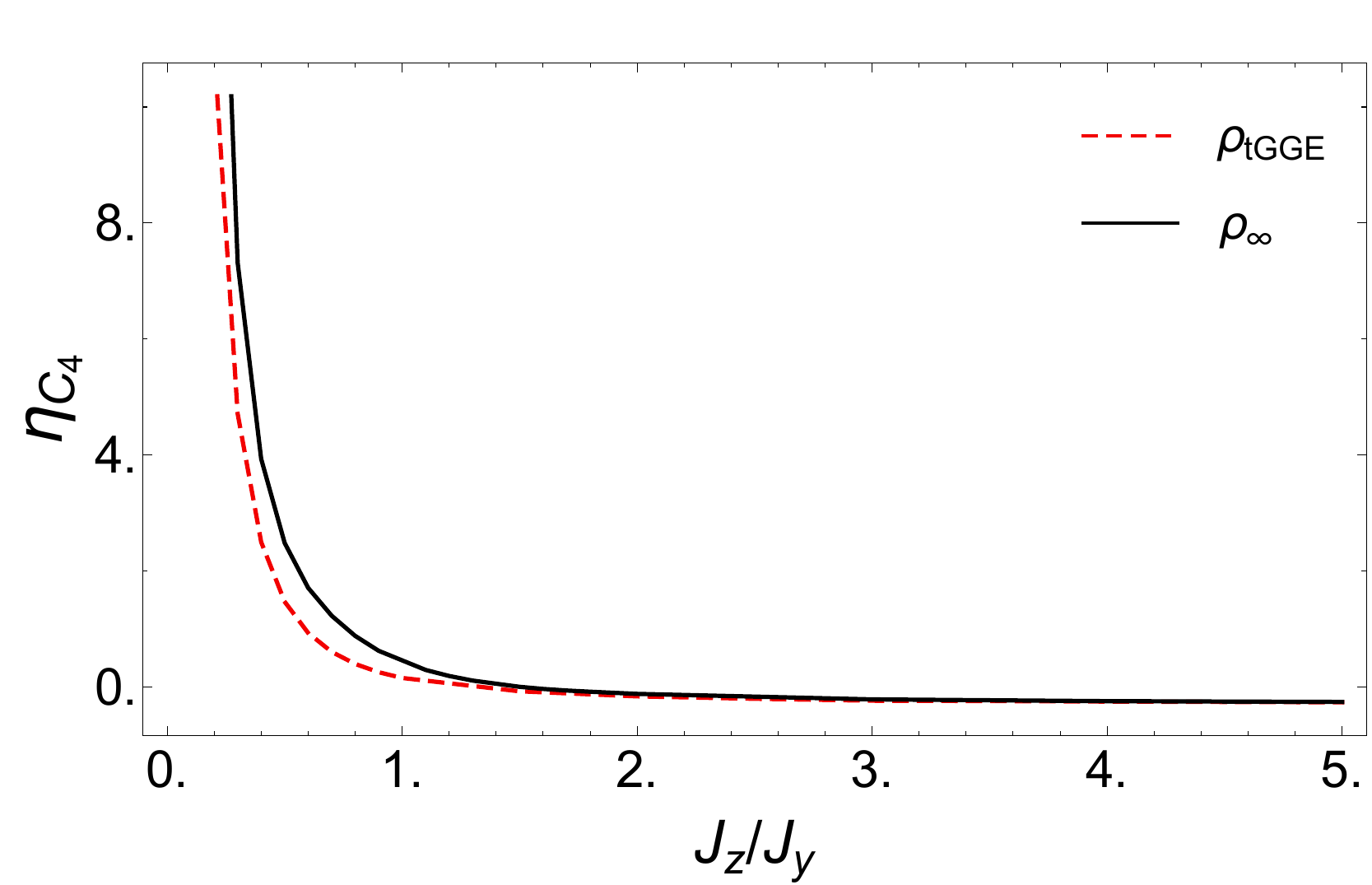}
\caption{Ratio $\eta_{C_4}$, Eq.~\eqref{EqEta}, measures how non-thermal the steady state is at different anisotropies $J_z/J_y$ for $J_y=h=1$, $\gamma=0.5$. 
This result suggests that the experiment should optimally be performed at small $J_z/J_y$ or $J_z=0$.
The calculation is based on the exact steady state $\rho_{\infty}$ at $\epsilon=0.01$ on $N=6$ sites and on the truncated GGE $\rho_{\textrm{tGGE}}$ on $N=10$ sites using $N_C=4$ conservation laws. 
\label{FigNonTh} }
\end{figure}

\begin{figure}[b!]
\center
\includegraphics[width=.9\linewidth]{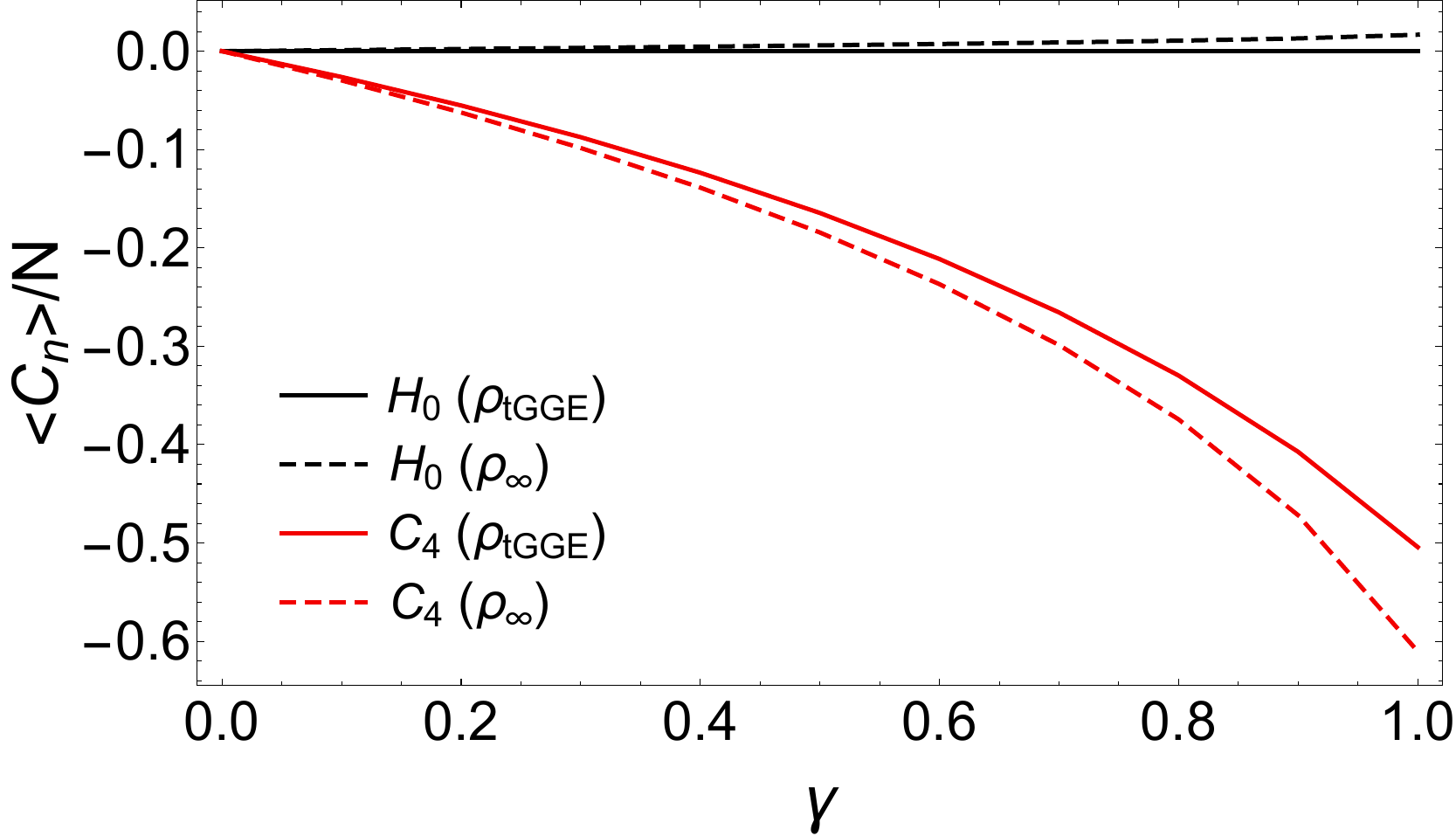}
\caption{Expectation  values  $\ave{H_0}$ and $\ave{C_4}$ as a function of relative driving strength $\gamma$, Eqs.~(\ref{EqLindZ1},\ref{EqLindZ2}), for $J_z=0$, corresponding to the transverse-field Ising case. Stabilized steady state is obviously highly non-thermal and described with a GGE with $\beta\approx0$ and large $\lambda_4$, manifested via tiny $\ave{H_0}$ and large $\ave{C_4}$. Parameters: $J_y=h=1$, $J_z=0$; $N=10, N_c=4$ for $\rho_\textrm{tGGE}$ and $N=6, \epsilon=0.01$ for $\rho_\infty$.
\label{FigtIsing} }
\end{figure}

\begin{figure}[t!]
\center
\includegraphics[width=.9\linewidth]{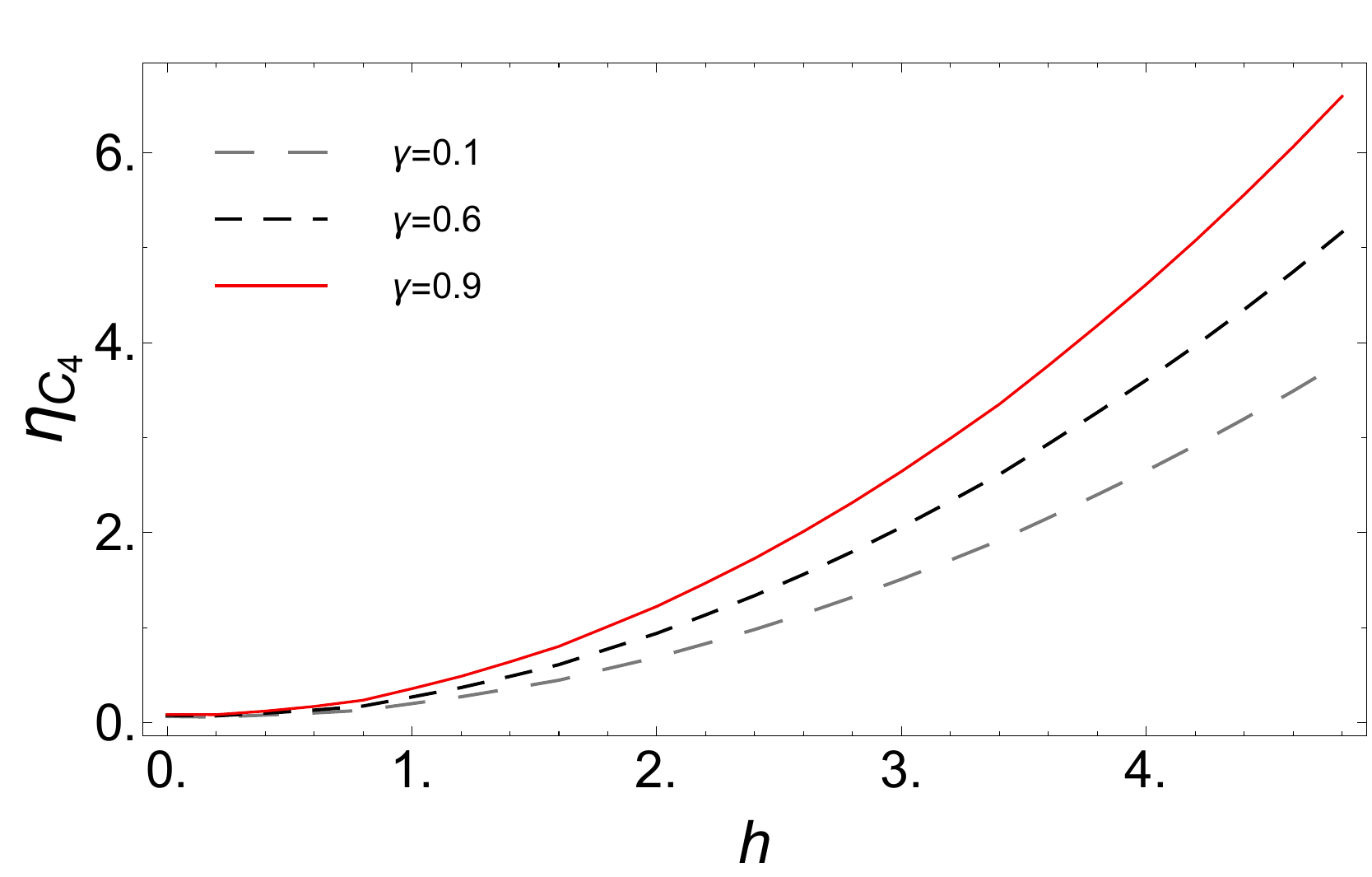}
\caption{Ratio $\eta_{C_4}$ as a function of magnetic field $h$ for weak anisotropy $J_z/J_y=0.9$ obtained from tGGE at $N=10$, $J_y=1$ and different ratios of Lindblad drivings $\gamma=0.1,0.6,0.9$. Increasing the magnetic field assists in preparing a more non-thermal state.
\label{FigNonThH} }
\end{figure}

In the following we focus on the operator $O=C_4$.
In the absence of Lindblad driving $\cl{D}^{1(2)}$, the ratio equals $\eta_{C_4}=0$ due to integrability breaking power-law decay of interactions in the Hamiltonian. In the presence of a weak Lindblad drive, on the other hand, the steady state can be highly non-thermal.
Figure \ref{FigNonTh} shows $\eta_{C_4}$ as a function of anisotropy $J_z/J_y$, obtained from the exact density matrix $\rho_\infty$ on $N=6$ sites at $\epsilon=0.01$ and from a $\rho_{\textrm{tGGE}}$ on $N=10$. 
The dependence on $J_z/J_y$ suggests that the experiment observing a highly non-thermal steady state ($\eta_{C_4}\gg 1$) achieved by a weak driving should operate at a small $J_z/J_y$ or even at $J_z=0$ (corresponding to the transverse-field Ising case). 

Why $\eta_{C_4}$ is so large for $J_z=0$ (and small $J_z$) can be reasoned by looking directly at the expectation values of $\ave{H_0}$ and $\ave{C_4}$, Fig.~\ref{FigtIsing}. $\ave{H_0}\approx 0$ is almost zero, suggesting an almost infinite temperature state ($\beta\approx0$), which would imply $\ave{C_4}\approx 0$. On the other hand, the observed  $\ave{C_4}\sim 1$ is actually large. This observation, together with a good agreement between $\rho_\infty$ and $\rho_{\textrm{tGGE}}$, clearly shows that a GGE with a large $\lambda_4$ is stabilized, despite $\beta\approx0$. Measuring $\ave{H_0}$ and $\ave{C_4}$ would thus serve as strong affirmation of our theory. 

In Fig.~\ref{FigNonThH}, we show that at mild anisotropy, for example, $J_z/J_y=0.9$, also a magnetic field $h$ helps to prepare a more non-thermal state.

\section{Dissipation engineering}
\label{SecImp}

Having shown that our dissipative driving can stabilize a steady state described by a GGE, we now discuss the implementation of the desired dynamics in a trapped-ion setup.
To engineer suitable dissipative interactions, we combine coherent couplings with sources of dissipation such as induced spontaneous emission and sympathetic cooling. We use these tools to engineer the desired one- and two-body jump operators and verify their action numerically.

In this section, we assume that the YZ-Hamiltonian in Eq. \eqref{EqYZh} is implemented. This allows us to engineer the dissipation in Eqs. \eqref{EqLindZ1}--\eqref{EqLindZ2} between the levels $\ket{\up}, \ket{\dn}$. As these are eigenstates of (stimulated) spontaneous emission, it is possible to use repumper beams as sources of dissipation.
In Sec. \ref{sec:generalization}, in turn, we assume that the XY-model in Eq.~\eqref{EqXYh} is realized. This requires the dissipation to be engineered between the eigenstates of $\sigma_x$, as is demonstrated using sympathetic cooling.

\begin{figure}[t]
\centering
\includegraphics[width=\columnwidth]{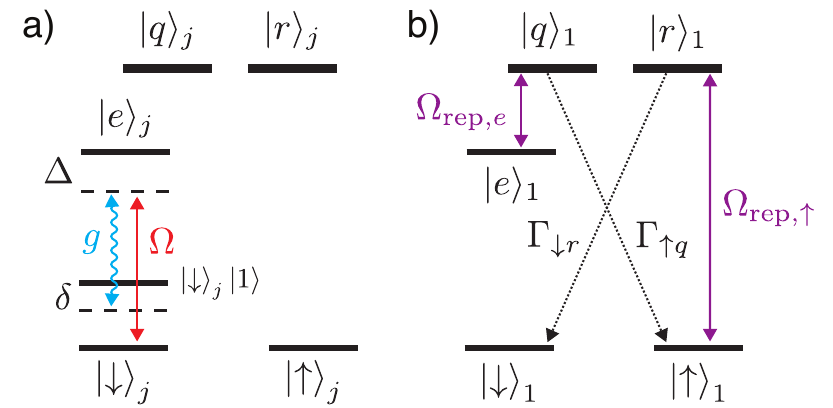}
\caption{
Setup for engineered single-body and two-body dissipation. We consider two trapped ions, $j=1$ and $j=2$, with two stable ground levels, $\ket{\dn}$ and $\ket{\up}$, and three excited levels, $\ket{e}$, $\ket{r}$, and $\ket{q}$. The ions are subject to coherent couplings (solid lines) and dissipation (dotted lines). (a) A weak drive excites both ions from $\ket{\dn}$ to $\ket{e}$ (strength $\Omega$, detuning $\Delta$).
The transition from $\ket{e}$ to $\ket{\dn}$ is coupled to the motional mode $a$ (phonon detuning $\delta$) by a red-sideband interaction (coupling constant $g$). The second ket denotes motional excitation and is dropped when in vacuum.
(b) We realize single-body decay from $\ket{\up}$ to $\ket{\dn}$ and from $\ket{e}$ to $\ket{\up}$ on ion $1$. This is done by optical pumping using tunable repumper beams (Rabi rates $\Omega_{\mathrm{rep},\up}$ and $\Omega_{\mathrm{rep},e}$), via the unstable levels $\ket{r}$ and $\ket{q}$ (decay rates $\Gamma_{\dn r}$ and $\Gamma_{e \up}$).
}
\label{fig:setup}
\end{figure}

\subsection{Setup}
\label{sec:setup}

To implement the one- and two-body jump operators in Eqs. \eqref{EqLindZ1} and \eqref{EqLindZ2}, we consider a system of trapped ions coupled through motional modes.
To simplify the discussion, we start by considering a minimal instance consisting of two ions indexed $1$ and $2$, and a motional mode $a$, and generalize to more ions in Sec. \ref{sec:scalability}--\ref{sec:micro}.
As is shown in Fig. \ref{fig:setup}, each of the ions is assumed to have two stable ground levels, $\ket{\uparrow}$ and $\ket{\downarrow}$, and three excited levels, $\ket{e}$, $\ket{r}$, and $\ket{q}$. The motional mode $a$ is assumed to be cooled to the ground state, $\ket{0}$. The free Hamiltonian of this system is given by
\begin{align}
H_\mathrm{free} = \delta a^\dagger a + \sum_{j=1}^2 \Delta \ket{e}_j \bra{e} + \Delta_r \ket{r}_j \bra{r} + \Delta_q \ket{q}_j \bra{q}.
\label{eq:Hfree}
\end{align}
Here we introduce a phonon detuning $\delta$ and an ionic detuning $\Delta$, assuming that we work in a suitable rotating frame with respect to the fields to be introduced below.
We will use level $\ket{r}$ to realize the single-body decay in Eq.~\eqref{EqLindZ1} and levels $\ket{e}$ and $\ket{q}$ in combination with mode $a$ for the two-body dissipation in Eq.~\eqref{EqLindZ2}. Level $\ket{q}$ will be used to add a dissipation channel from level $\ket{e}$ to level $\ket{\up}$.
The ions are excited from $\ket{\dn}$ to $\ket{e}$ using a weak ``carrier'' drive
\begin{align}
H_\mathrm{drive} = \frac{\Omega}{2} \sum_{j=1}^2 \ket{e}_{j} \bra{\dn} + \mathrm{H.c.},
\label{eq:Hdrive}
\end{align}
with a Rabi frequency $\Omega$. In addition, levels $\ket{\up}$ and $\ket{e}$ of ion $1$ are excited to $\ket{r}$ and $\ket{q}$ by coherent ``repumper'' beams,
\begin{align}
H_{\mathrm{rep},\up} &= \frac{\Omega_{\mathrm{rep},\up}}{2} \ket{r}_{1} \bra{\up} + \mathrm{H.c.}
\label{eq:Hrep}
\\
H_{\mathrm{rep},e} &= \frac{\Omega_{\mathrm{rep},e}}{2} \ket{q}_{1} \bra{e} + \mathrm{H.c.}
\label{eq:Hrep:e}
\end{align}
with Rabi rates $\Omega_{\mathrm{rep},\up/e}$.
The coupling between ions $1$ and $2$, needed to engineer two-body dissipation, is mediated by a common motional mode, with creation (annihilation) operator $a^\dagger$ ($a$). This phonon mode is coupled to the transition from $\ket{e}$ to $\ket{\dn}$ by the red-sideband interaction
\begin{align}
H_\mathrm{int} = g \sum_{j=1}^2 a^\dagger \ket{\dn}_{j} \bra{e} + a \ket{e}_{j} \bra{\dn} + \mathrm{H.c.},
\label{eq:Hsideband}
\end{align}
with a coupling constant $g$.

In addition to the above coherent interactions, the excited level $\ket{r}$ is assumed to be inherently unstable and to decay to level $\ket{\dn}$ by spontaneous emission, which can be described using the jump operators,
\begin{align}\label{EqL0r}
L_{\dn r,j} = \sqrt{\Gamma_{\dn r}} \ket{\dn}_j \bra{r}, ~~~ (j=1,2).
\end{align}
The excited level $\ket{q}$, in turn, is assumed to decay to $\ket{\up}$,
\begin{align}\label{EqL1q}
L_{\up q,j} = \sqrt{\Gamma_{\up q}} \ket{\up}_j \bra{q}, ~~~ (j=1,2).
\end{align}

To describe the joint dynamics of the ions and the phonons, we use the following notation: the state of the system is described by two kets, where the first ket denotes the internal state of the ions, e.g., $\ket{\dn \dn} = \ket{\dn}_{1}\ket{\dn}_{2}$. Motional excitations are denoted by a second ket, e.g., $\ket{\dn \dn} \ket{1}$, which is dropped when the motion is in the ground state.

\subsection{Single-body dissipation}
\label{sec:repump}
To realize single-body dissipation, we employ standard optical pumping, combining excitation from $\ket{\up}$ to $\ket{r}$ by $H_{\mathrm{rep},\up}$, Eq.~\eqref{eq:Hrep}, and decay from $\ket{r}$ to $\ket{\dn}$ by spontaneous emission $L_{\dn r,j}$. The effective jump operator \cite{Reiter12} for the decay of level $\ket{\up}$ to $\ket{\dn}$ through $\ket{r}$ is thus, after elimination of level $\ket{r}$, given by
\begin{align}
L_\mathrm{eff}^{(1)}
= \sqrt{\gamma_1} \ket{\dn}_1 \bra{\up} 
\equiv \sqrt{ \frac{\Omega_{\mathrm{rep},\up}^2}{\Gamma_{\dn r}} } \ket{\dn}_1 \bra{\up},
\label{eq:Lrep}
\end{align}
We thereby realize the desired jump operator $L_1^{(1)}=\sqrt{\epsilon(1-\gamma)}\, S^-_1$ in Eq.~\eqref{EqLindZ1}.
Using individual addressing techniques, this process can be made site-specific.
The decay rate $\gamma_1$ can be tuned by varying $\Omega_{\mathrm{rep},\up}$, assuming it to be much smaller than the natural linewidth of level $\ket{r}$, $\Gamma_{\dn r} \gg \Omega_{\mathrm{rep},\up}$.
Note that, while here we have only assumed the desired dissipation channel from $\ket{r}$ to $\ket{\dn}$, additional decay processes from these levels could be described by the same method.

To engineer the two-body dissipation in Eq.~\eqref{EqLindZ2} in Sec. \ref{sec:mechanism} below, we will also rely on an induced spontaneous emission process from $\ket{e}$ to $\ket{\up}$. Here we assume that we can realize the YZ-Hamiltonian in Eq. \eqref{EqYZh}. Alternatively, in the presence of the XY-Hamiltonian in Eq. \eqref{EqXYh}, sympathetic cooling can be used as a source of dissipation, as is described in Sec. \ref{sec:generalization}.

To realize optical pumping from $\ket{e}$ to $\ket{\up}$, we couple $\ket{e}$ of ion $1$ to the unstable level $\ket{q}$ (total decay rate $\Gamma_q$) using the repumper in Eq. \eqref{eq:Hrep:e}. Together with the decay from $\ket{q}$ to $\ket{\up}$ in Eq. \eqref{EqL1q}, this realizes
\begin{eqnarray}
L_{\up e}
= \sqrt{\Gamma} \ket{\up}_1 \bra{e}
\equiv \sqrt{ \frac{\Omega_{\mathrm{rep},e}^2}{\Gamma_{\up q}} } \ket{\up}_1 \bra{e}.
\label{eq:Lrep:e}
\end{eqnarray}
Again, the decay rate $\Gamma$ is tunable through the strength of the corresponding repumper beam $\Omega_{\mathrm{rep},e}$.

\begin{figure}[t]
\centering
\includegraphics[width=\columnwidth]{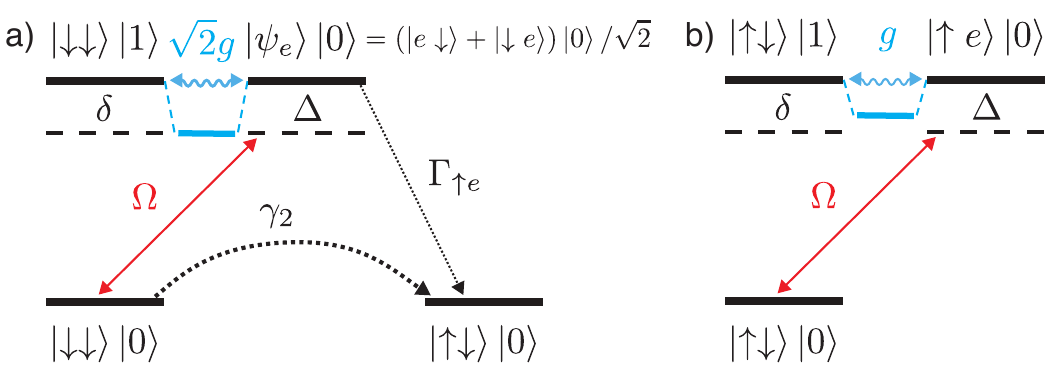}
\caption{
Engineering of the two-body dissipation.
(a) Desired effective decay process. State $\ket{\dn\dn} \ket{0}$ is coupled to the ion-excited state $\ket{\psi_e}\ket{0} = \frac{1}{\sqrt{2}}(\ket{e \dn} + \ket{\dn e})\ket{0}$ by the drive $\Omega$. $\ket{\psi_e}\ket{0}$ is strongly coupled to the motion-excited states $\ket{\dn\dn} \ket{1}$ by the sideband coupling $g$, which is enhanced by a factor $\sqrt{2}$ due to constructive interference. For $\Delta \delta = 2 g^2$, the lower dressed state of $\ket{\psi_e}\ket{0}$ and $\ket{\dn\dn}\ket{1}$ (indicated in blue) is in resonance with the drive and hence rapidly excited from $\ket{\dn\dn}\ket{0}$. Through its contribution from $\ket{\psi_e}\ket{0}$, it decays to $\ket{\up\dn} \ket{0}$ by spontaneous emission $\Gamma$. These resonant couplings form an effective decay process from $\ket{\dn\dn} \ket{0}$ to $\ket{\up\dn} \ket{0}$, at an enhanced rate $\gamma_2$.
(b) Undesired process. Also state $\ket{\up\dn} \ket{0}$ is excited by the drive, to an ion-excited state $\ket{e \up}\ket{0}$. The sideband coupling couples $\ket{e \up}\ket{0}$ to $\ket{\up\dn}\ket{1}$ at a coupling constant $g$. Their dressed states are thus out of resonance and only weakly populated by the drive.
}
\label{fig:mechanism}
\end{figure}

\subsection{Two-body dissipation}
\label{sec:mechanism}

We now turn to the two-body dissipation
in Eq.~\eqref{EqLindZ2}. 
Compared to the single-body dissipation in the previous section, the operator $L_j^{(2)}$ is more complicated to engineer, but itself sufficient to realize a highly non-thermal GGE (see Fig.~\ref{FigExpVal} at $\gamma=1$).
For our minimal instance of two ions, the operator reads
$L^{(2)}_1 = \sqrt{\epsilon\gamma}\,S_1^+ P_{2}^{\downarrow} = \sqrt{\epsilon\gamma}\, \ket{\up \dn} \bra{\dn \dn}$.
The action of this operator can be understood as a raising on spin $1$, $S_1^+ = \ket{\up}_1 \bra{\dn}$, conditioned on the state of spin $2$.

We will now engineer the desired two-body dissipation based on the assumptions of weak driving, $\Omega^2 \ll \{\Gamma^2, g^2\}$, and strong coupling, $\Gamma^2 \ll g^2$.
In this regime, the ground state $\ket{\dn \dn} \ket{0}$ is weakly excited by $H_\mathrm{drive}$ to the excited state $\ket{\psi_e} \ket{0} = \frac{1}{\sqrt{2}}( \ket{e \dn} + \ket{\dn e} ) \ket{0}$, which comprises of a superposition of excitations of both ions. Further excitation to the double-excited state $\ket{ee}$ can be neglected, as we will see further down. We engineer the desired mechanism using the couplings of $\ket{\psi_e} \ket{0}$:

The ion-excited state $\ket{\psi_e} \ket{0}$ is coupled to the motion-excited state $\ket{\dn \dn} \ket{1}$ by the red-sideband coupling $H_\mathrm{int}$. Due to constructive interference between the excitation of the two ions, the corresponding rate is given by $\sqrt{2} g$. The Hamiltonian for the coupled subspace is
\begin{align}
\label{eq:He00}
H_\mathrm{e,\dn\dn}
&= \Delta \ket{\psi_e} \ket{0} \bra{0} \bra{\psi_e} + \delta \ket{\dn \dn} \ket{1} \bra{1} \bra{\dn \dn}
\\ \nonumber
&+ \sqrt{2} g ( \ket{\psi_e} \ket{0} \bra{1} \bra{\dn \dn} + \ket{\dn \dn} \ket{1} \bra{0} \bra{\psi_e}),
\end{align}
and illustrated in Fig.~\ref{fig:mechanism} a).
Based on the assumption of a weak drive compared to the rapid dynamics of the excited subspace, we make a separation of timescales and first regard $H_\mathrm{e,\dn\dn}$ alone, without the drive:

Provided strong coupling, the excited states of the strongly coupled subspace hybridize and form dressed states $\ket{\psi_\pm}$ at detunings
\begin{eqnarray}
\Delta_\pm = \frac{\Delta + \delta}{2} \pm \frac{1}{2} \sqrt{(\Delta + \delta)^2 + 4 ( \Delta \delta - 2 g^2) }.
\end{eqnarray}
Setting the ionic and the motional detunings to $\Delta \delta = 2 g^2$ (e.g., $\Delta = \delta = \sqrt{2} g$) brings the lower dressed state in resonance with the drive $\Omega$, i.e., $\Delta_- = 0$. As a consequence, $\ket{\dn\dn}$ is resonantly excited to $\ket{\psi_e}$ which in turn decays to $\ket{\up\dn}$ at a rate $\Gamma$. This results in the required effective decay from $\ket{\dn\dn}$ to $\ket{\up\dn}$, mediated by the resonant lower dressed state, $\ket{\psi_-}$. 
In App.~\ref{sec:analysis} we present a full microscopic derivation that yields the two-body Lindblad operator, Eq.~\eqref{EqLindZ2},
\begin{align}
\label{eq:LeffGamma}
L^{(2)}_\textrm{eff} = \sqrt{\gamma_2} \ket{\up\dn}_{1,2}\bra{\dn\dn}.
\end{align}
Using second-order perturbation theory in $\Omega/\Gamma$ the effective decay rate is given by,
\begin{align}
\gamma_2 &= \frac{4 \Omega^2}{\Gamma}.
\label{eq:gammaplusPB}
\end{align}
The decay rate $\gamma_2$, and hence the relative strength of the single- and two-body dissipation, can thus be adjusted by varying $\Omega$ and $\Gamma(\Omega_{\mathrm{rep},e})$.
$\gamma_2$ is ultimately limited by the linewidth $\Gamma_q$ of level $\ket{q}$, which is involved in constructing the engineered single-body decay in Eq.~\eqref{eq:Lrep:e}.

Compared to $\gamma_2$, effective decay processes mediated by $\ket{ee}$ can be neglected: $\ket{ee}$ forms a coupled two-excitation subspace with $\ket{\psi_e}\ket{1}$ and $\ket{\dn\dn}\ket{2}$ (couplings $\sqrt{2}g$ and $2g$). While for the above parameter choice, the drive from $\ket{\psi_e}$ is in resonance with the two-excitation dressed states, the coupling rates of the so mediated effective process only enter to fourth order in perturbation theory and are thus negligibly small.
Also AC Stark shifts arising from the weak off-resonant excitation of $\ket{\up\dn}$ (cf. App.~\ref{sec:ACStark}) can be safely ignored in the considered parameter regime.

Another imperfection inherent to the scheme is given by the population of the excited level $\ket{\up e}$ which is off-resonantly excited from $\ket{\up\dn}$ by the drive $\Omega$, as illustrated in Fig. \ref{fig:mechanism} b). For perfect individual addressing of the first ion by $H_{\mathrm{rep},e}$, $\ket{\up e}$ is steadily populated. Using adiabatic elimination, it is possible to estimate the steady-state population of an excited state such as $\ket{\up e}$ \cite{Reiter16, Reiter17}, which scales
\begin{align}
P_{\up e} \sim \frac{\Omega^2}{g^2}.
\end{align}
Population of the excited state $\ket{\up e}$ is thus expected to be largely suppressed, as we numerically confirm in the following.

\begin{figure}[t]
\centering
\includegraphics[width=\columnwidth]{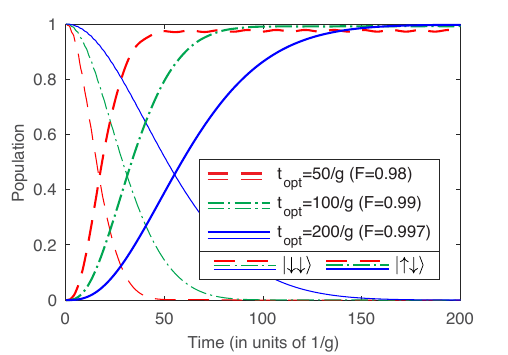}
\caption{
Numerical simulation. We verify the engineering of the two-body dissipation by simulating the dynamics and plotting the population of the states $\ket{\dn\dn}$ (thin lines) and $\ket{\up\dn}$ (thick lines) over time, starting from an initial state $\ket{\dn\dn}$. We present three different parameter choices, where we optimize the action of the scheme at different times $t_\mathrm{opt} = \{ 50, 100, 200\}/g$ (dashed lines, dash-dotted lines, solid lines). We obtain fidelities $F_\mathrm{opt}=\{0.98, 0.99, 0.997\}$ for the parameter choices $\Gamma_\mathrm{opt} = \{0.82, 0.48, 0.29 \}g$ and $\Omega_\mathrm{opt} = \{0.15, 0.08, 0.05 \}g \approx \Gamma_\mathrm{opt}/6$. Population in the excited states is negligible.
}
\label{fig:simulation}
\end{figure}

We verify the engineering of the two-body dissipation and assess its performance numerically by simulating the dynamics given by the master equation
\begin{align}
\dot{\rho} &= -i [H_\mathrm{free} + H_\mathrm{drive} + H_\mathrm{int},\rho] +  \mathcal{D}^{(\up e)}\rho,\\
\mathcal{D}^{(\up e)}\rho
&= L_{\up e} \rho L_{\up e}^\dagger - \frac{1}{2}\{L_{\up e}^\dagger L_{\up e}, \rho\} \notag
\end{align}
We use this to obtain i) the optimal parameter choice, ii) the extent of unwanted population of the excited level $\ket{\up e}$, and iii) the timescales of the dissipative compared to the unitary dynamics.

We assume that the system starts from $\ket{\dn\dn} \ket{0}$ and optimize the fidelity of the state $\ket{\up\dn} \ket{0}$ after a chosen time, $t_\mathrm{opt} = \{ 50, 100, 200 \}/g$ by the choice of the parameters $\Gamma$ and $\Omega$.
The result is plotted in Fig. \ref{fig:simulation}. From the initial state $\ket{\dn\dn}$, the system evolves to fidelities with $\ket{\up\dn}$ of $F_\mathrm{opt}=\{0.98, 0.99, 0.997\}$ for the parameter choices $\Gamma_\mathrm{opt} = \{0.82, 0.48, 0.29 \}g$ and $\Omega_\mathrm{opt} = \{0.15, 0.08, 0.05 \}g \approx \Gamma_\mathrm{opt}/6$.
After a short non-exponential transient ($t < 1/\gamma_\mathrm{eff}$) which builds up the excited state population, the curves attain the desired exponential form that is described by the effective jump operator in Eq. \eqref{eq:LeffGamma}. This represents a close approximation to the desired dynamics.

The optimal parameters fulfill the conditions for weak driving and strong coupling, $\Omega^2 \ll \Gamma^2 \ll g^2$, which are the assumptions used in the above analysis. The residual population in $\ket{\up e}$ is found to be $P_\mathrm{\up e} \approx \{ 0.02, 0.008, 0.003\}$ and thus indeed negligible.

For the effective dissipation rate in Eq. \eqref{eq:gammaplusPB} we obtain $\gamma_2 = \{0.054, 0.027, 0.014\}g$. For typical values of $g/(2\pi) \sim 10 ~ \mathrm{kHz}$, this yields $ \gamma_2 = \{3.4, 1.7, 0.89\}/\mathrm{ms}$. The convergence time $\tau = 1/\gamma_2$ (corresponding to the decay of the initial-state population to 1/e) is found to be $ \tau = \{18, 37, 70\}/g = \{0.29, 0.59, 1.1\} ~\mathrm{ms}$, in good agreement with the results plotted in Fig. \ref{fig:simulation}.

The achievable values for $\gamma_2$ should be compared with the typical coupling strength in realizations of the spin models of about $J_{x/y}/(2\pi) \sim 100 ~ \mathrm{Hz}$ \cite{Jurcevic2014, Richerme2014}, and the corresponding timescale of $\tau_\mathrm{spin} \sim 1 ~\mathrm{m s}$.
From the above numbers it can be seen that the effective dissipation rate can be tuned to values within the same order of magnitude as the coupling constants of the spin model. Obtaining smaller values for $\gamma_2$ -- and thus making the dissipation into a perturbation as assumed in Sec. \ref{sec::Model}--\ref{sec::NESS} -- is in turn achieved by choosing weaker repump and driving rates, $\Gamma$ and $\Omega$.

\begin{figure}[t]
\centering
\includegraphics[width=\columnwidth]{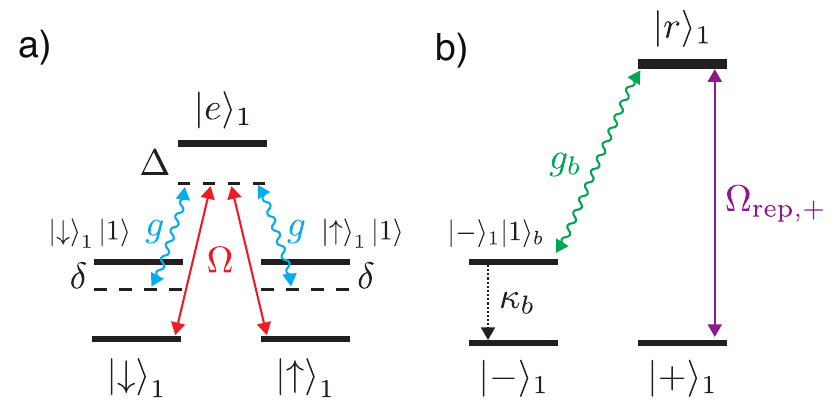}
\caption{
Generalization of the couplings to the $x$-basis. (a) To engineer two-body dissipation in the $x$-basis, we use couplings between the states $\ket{\pm} = ( \ket{\dn} \pm \ket{\up} ) / \sqrt{2}$, and $\ket{e}$. This is achieved by coupling the levels $\ket{\dn}$ and $\ket{\up}$ coherently to $\ket{e}$.
(b) Dissipation in the $x$-basis is facilitated by sympathetic cooling of the motion. The single-body decay process from $\ket{+}$ to $\ket{-}$ is engineered by excitation from $\ket{+}$ by a weak drive ($\Omega_\mathrm{rep, +}$) to an auxiliary level $\ket{r}$, a sideband coupling of the transition $\ket{r} \rightarrow \ket{-}$ to a motional mode $b$ ($g_b$), and sympathetic cooling of $b$ ($\kappa_b$). Decay from $\ket{e}$ to $\ket{+}$ (not shown) is engineered accordingly using a motional mode $c$, as is described in the text.
}
\label{fig:setup_x}
\end{figure}

\section{Generalization of the couplings to the $x$-basis}
\label{sec:generalization}

Spin models along arbitrary directions, with and without anisotropy, can be realized on trapped-ion platforms, such as Paul traps and Penning traps, and also in microtraps \cite{Porras2004}.
The majority of the available setups, however, support XY-Hamiltonians without anisotropy \cite{Jurcevic2014}. In the preceding section, we assumed the less common YZ spin Hamiltonian which enabled us to engineer the dissipation in the $z$-basis, the eigenbasis of decay by spontaneous emission.
As an alternative to such implementation, we can utilize the more standard XY-Hamiltonian in Eq. \eqref{EqXYh}. We should note, however, that in order to stabilize a non-trivial ($\rho_\infty\neq \mathbb{1}$) and distinguishingly non-thermal steady state, an engineered form of Lindblad operators with proper symmetries must be used. For example, the XY-Hamiltonian in combination with the dissipation in the $z$-basis would not result in a large $\ave{C_4}$. This problem is resolved by using the XY-Hamiltonian in combination with dissipation in the rotated $x$-basis, the engineering of which is presented below.

We point out that a simpler alternative is given by using the XY-Hamiltonian in combination with {\it any} type of Lindblad dissipation. Such setting would yield a non-trivial dynamics towards a possibly trivial steady state, where {\it time evolution} is approximately captured with a time-dependent GGE \cite{lange18}. However, the non-thermal features might not be very pronounced in a generic setup, therefore we focus here on a setup with non-trivial steady states with clearly non-thermal nature.

We now engineer the dissipation in the $x$-direction which can stabilize a steady-state GGE,
\begin{align}
\label{EqLindX1}
&L_j^{(1,x)}=\sqrt{\epsilon(1-\gamma)}S_{j,x}^{-}.
\\
\label{EqLindX2}
&L_j^{(2,x)}=\sqrt{\epsilon\gamma}S_{j,x}^+ P_{j+1,x}^{\downarrow}, \quad \quad  
\end{align}
when combined with the XY Hamiltonian. Formally speaking, the change from the z- to the $x$-basis amounts to replacing the eigenstates of $\sigma_z$, $\{\ket{\dn},\ket{\up}\}$ with the eigenstates of $\sigma_x$, $\ket{\pm}=( \ket{\dn} \pm \ket{\up} )/\sqrt{2}$, i.e., $\ket{\dn} \mapsto \ket{-}$, $\ket{\up} \mapsto \ket{+}$, in all steps of our previous derivation performed for the dissipation in the $z$-direction.
Physically, using $\ket{\pm}$ is realized by coupling to both transitions $\ket{\dn} \leftrightarrow \ket{e}$ and $\ket{\up} \leftrightarrow \ket{e}$ coherently. The resulting interactions, illustrated in Fig. \ref{fig:setup_x} a), read
\begin{align}
H_\mathrm{drive, x} &= \frac{\Omega}{2} \sum_{j=1}^2 \ket{e}_{j} \bra{-} + \mathrm{H.c.},
\label{eq:Hdrive:x}
\\
H_\mathrm{int, x} &= g  \sum_{j=1}^2 \left(a^\dagger \ket{-}_{j} \bra{e} + a \ket{e}_{j} \bra{-} \right).
\label{eq:Hsideband:x}
\end{align}
On the other hand, decay by spontaneous emission, as utilized in the previous sections naturally occurs in the $z$-basis, $\{\ket{\dn},\ket{\up}\}$.
In contrast to this, we now need to engineer sources of dissipation in the $x$-basis, which replaces Eqs.~\eqref{EqL0r},\eqref{eq:Lrep:e} in the $z$-basis. 
In the following, we demonstrate how to achieve this using decay of excitations via the motional degree of freedom by sympathetic cooling.

As shown in Fig.~\ref{fig:setup_x} b), to implement the single-body decay in Eq.~\eqref{EqLindX1}, we excite $\ket{+}$ to the auxiliary level $\ket{r}$ by a repumper
\begin{align}
H_\mathrm{rep,+} = \Omega_\mathrm{rep, +} \ket{r}_1 \bra{+} + \mathrm{H.c.}
\end{align}
The excitation to level $\ket{r}$ is transferred coherently to an auxiliary motional mode $b$ using a sideband interaction,
\begin{align}
H_\mathrm{b} = g_\mathrm{b} b^\dagger \ket{-}_1 \bra{r} + \mathrm{H.c.},
\end{align}
with a coupling constant $g_\mathrm{b}$. Mode $b$ is subject to sympathetic cooling which realizes the jump operator
\begin{align}
L_\mathrm{b} = \sqrt{\kappa} b.
\end{align}
Adiabatic elimination of $b$ leads, for $g_b \ll \kappa$, to the desired decay channel
\begin{align}
L_{-r}=\sqrt{\Gamma_{-r}}\ket{-}_1\bra{r},
\end{align}
with a rate $\Gamma_{-r} = g_b^2/\kappa$, in analogy to Eq. \eqref{EqL0r}.
We realize the decay $L_{+e}=\sqrt{\Gamma_{+e}}\ket{+}_1\bra{e}$ (cf. Eq. \eqref{eq:Lrep:e}) in a similar fashion, utilizing a motional mode $c$, which is subject to sympathetic cooling. Here, we couple $\ket{e}\ket{0}_c$ to $\ket{+}\ket{1}_c$ by a sideband drive ($g_c$), which then decays to $\ket{+}\ket{0}_c$ by sympathetic cooling ($\kappa_c$), resulting in an effective decay rate $\Gamma_{+e} = g_c^2/\kappa_c$. Involving level $\ket{q}$ is not necessary.
Carrying out the same analysis as in Sec. \ref{sec:analysis}, we obtain the effective operator
\begin{align}
\label{eq:LeffGammax}
L_{\mathrm{eff}}^{(2,x)} = \sqrt{\gamma_\mathrm{2,x}} \ket{+-}_{1,2}\bra{--},
\end{align}
with a tunable decay rate $\gamma_\mathrm{2,x} = 4 \Omega^2 / \Gamma_{+e}^2$.
We have thus realized the desired two-body dissipation in the $x$-basis in Eq.~\eqref{EqLindX2} by means of sympathetic cooling.

\begin{figure*}[t]
\centering
\includegraphics[width=17.2cm]{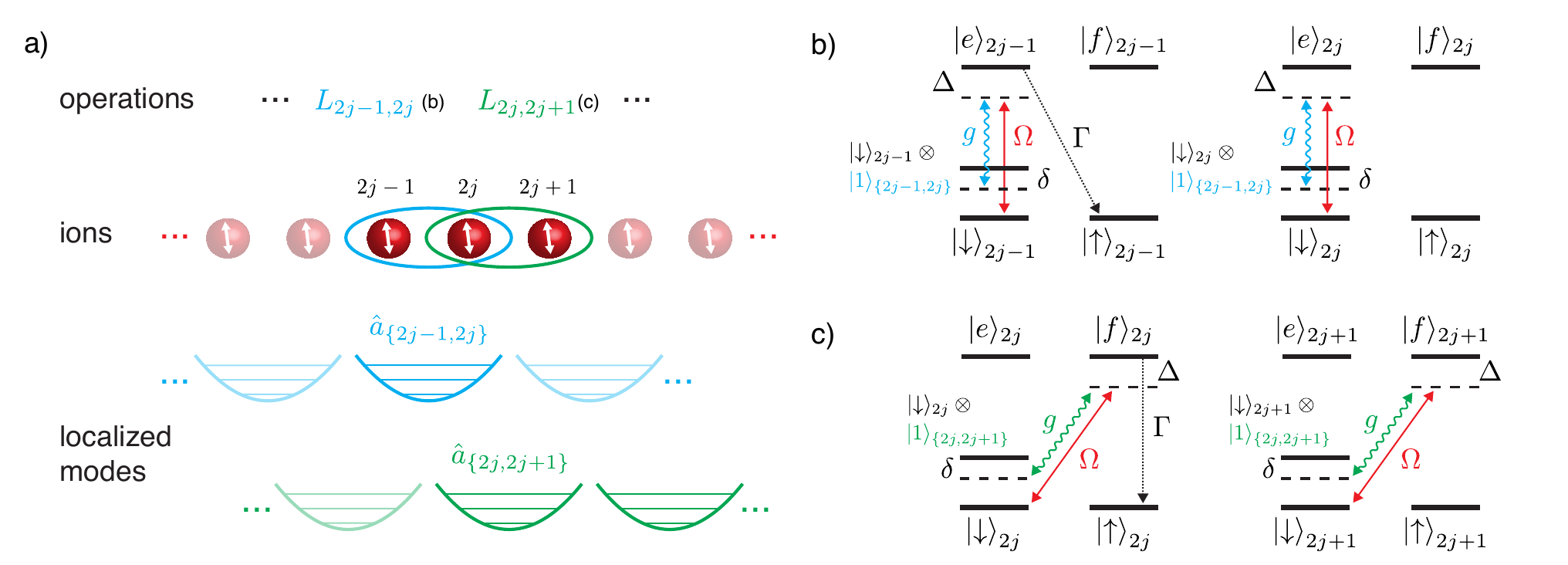}
\caption{
Scalable implementation in trapped ions. (a) Setup. We assume a string of $N$ ions, denoted $2j \pm 1$ for odd ions and $2j$ for even ions. Two-body dissipation represented by the jump operators $L_{2j-1,2j}$ and $L_{2j-1,2j}$ is realized by coupling pairs of ions to localized phonon modes. Two sets of motional modes, $a_{2j-1,2j}$ and $a_{2j,2j+1}$, couple to pairs of ions $\{2j-1,2j\}$ and $\{2j,2j+1\}$. This avoids interference on the overlap ion $2j$. (b)-(c) Coupling configurations for pairs of ions $\{2j-1,2j\}$ (b) and $\{2j,2j+1\}$ (c). We assume two addressable excited levels, $\ket{e}$ and $\ket{f}$, for each ion. For odd ions $2j-1$ (even ions $2j$), we facilitate a decay from level $\ket{e}$ ($\ket{f}$) to level $\ket{\up}$ at a rate $\Gamma$, as is described in the text. The coherent drive $\Omega$ and the sideband couplings $g$ act on the transition from $\ket{e}$ to $\ket{\dn}$ (from $\ket{f}$ to $\ket{\dn}$) on pairs of ions $\{2j-1,2j\}$ ($\{2j,2j+1\}$).
}
\label{fig:scalability}
\end{figure*}

\section{Scalable implementation}
\label{sec:scalability}

Next, we discuss how to scale the mechanisms discussed in Sec.~\ref{SecImp}-\ref{sec:generalization} to larger numbers of ions.
For a scalable implementation of our scheme, we assume a chain of $N$ ions (with even $N$) and a level structure similar to Sec. \ref{sec:setup}.
The physical system for the scalable implementation of two-body dissipation in Eq.~\eqref{EqLindZ2} is shown in Fig. \ref{fig:scalability}. The implementation of such system based on ion micro-traps is presented in Sec. \ref{sec:micro}.
Here we seek to implement interactions on all pairs of ions, such as $\{2j-1, 2j\}$ and $\{2j, 2j+1\}$, as illustrated in Fig.~\ref{fig:scalability} a).
However, care has to be taken to avoid interference effects of the coherent couplings in the overlapping region, i.e., here ion $2j$.
We achieve this by devising two independent coupling configurations to mediate the engineered decay on the two different groups of ions, $\{2j-1,2j\}$ and $\{2j,2j+1\}$, as can be seen from Fig. \ref{fig:scalability} b)-c).
We assume each ion to have two (meta-) stable excited levels, $\ket{e}$ and $\ket{f}$, which are selectively addressable using, e.g., polarization selection rules.

For dissipation on pairs $\{2j-1, 2j\}$, level $\ket{e}$ is used to mediate the two-body dissipation, whereas for pairs $\{2j, 2j+1\}$ this is facilitated by level $\ket{f}$.
Correspondingly, we employ two sets of localized phonon modes: Modes $a_{2j-1,2j}$ interact with ions $\{2j-1,2j\}$, and modes $a_{2j,2j+1}$, couple to pairs $\{2j,2j+1\}$. The engineering of the mode structure will be discussed in detail in Sec. \ref{sec:micro}.

To implement two-body dissipation, we again use tunable optical pumping of $\ket{e}$ and $\ket{f}$ to $\ket{\up}$, in analogy to Sec. \ref{sec:mechanism}, Eq.~\eqref{eq:Lrep:e}. Utilizing different individually addressed repumper beams for ``odd'' ions $2j-1$ and ``even'' ions $2j$, we realize
\begin{eqnarray}
L_{\up e,2j-1}=&\sqrt{\Gamma} \ket{\up}_{2j-1} \bra{e},
\\
L_{\up f,2j}=&\sqrt{\Gamma} \ket{\up}_{2j} \bra{f}.
\end{eqnarray}
Odd ions $2j-1$ thus decay from $\ket{e}$ to $\ket{\up}$, whereas even ions $2j$ decay from $\ket{f}$ to $\ket{\up}$, both at an equal rate $\Gamma$.

For the continuous and measurement-free interrogation of the system, we use two sets of coherent drives,
\begin{align}
H_\mathrm{drive} &= H_{\mathrm{drive},e} + H_{\mathrm{drive},f}
\label{eq:Hdrive:scale1}
\\
H_{\mathrm{drive},e} &= \frac{\Omega}{2} \sum_{j=1}^{N/2} \left( \ket{e}_{2j-1} \bra{\dn} + \ket{e}_{2j} \bra{\dn} \right) + \mathrm{H.c.},
\\
H_{\mathrm{drive},f} &= \frac{\Omega}{2} \sum_{j=1}^{N/2} \left( \ket{f}_{2j} \bra{\dn} + \ket{f}_{2j+1} \bra{\dn} \right) + \mathrm{H.c.},
\label{eq:Hdrive:scale3}
\end{align}
coupling the ground level $\ket{\dn}$ to the excited level $\ket{e}$ or $\ket{f}$, as well as sideband interactions,
\begin{align}
H_\mathrm{int} &= H_{\mathrm{int},e} + H_{\mathrm{int},f},
\label{eq:Hsideband:scale1}
\\
H_{\mathrm{int},e} &= g \sum_{j=1}^{N/2} a^\dagger_{2j-1,2j} \left( \ket{\dn}_{2j-1} \bra{e} + \ket{\dn}_{2j} \bra{e} \right) + \mathrm{H.c.},
\\
H_{\mathrm{int},f} &= g \sum_{j=1}^{N/2} a^\dagger_{2j,2j+1} \left( \ket{\dn}_{2j} \bra{f} + \ket{\dn}_{2j+1} \bra{f} \right) + \mathrm{H.c.}
\label{eq:Hsideband:scale3}
\end{align}
These realize coupling configurations, by which the transition $\ket{e} \leftrightarrow \ket{\dn}$ ($\ket{f} \leftrightarrow \ket{\dn}$) of any pair of ions $\{2j-1, 2j\}$ ($\{2j-1, 2j\}$) is coupled to a localized motional mode $a_{2j-1,2j}$ ($a_{2j,2j-1}$).

As a result, following the recipe in Sec.~\ref{sec:mechanism}, we realize jump operators acting on pairs of ions over the whole chain,
\begin{align}
L_{2j-1}^{(2)} = \sqrt{\gamma_2} \ket{\up\dn}_{2j-1,2j} \bra{\dn\dn} = \sqrt{\gamma_2} S_{2j-1}^+ P_{2j}^{\downarrow} ,
\label{eq::EqLindZ2:scale:1}
\\
L_{2j}^{(2)} = \sqrt{\gamma_2} \ket{\up\dn}_{2j,2j+1} \bra{\dn\dn} = \sqrt{\gamma_2} S_{2j}^+ P_{2j+1}^{\downarrow}.
\label{eq::EqLindZ2:scale:2}
\end{align}
In the second step, these operators are brought back into the form of Eq.~\eqref{EqLindZ2}. Making the association $\gamma_2 = \epsilon \gamma$, we have thereby engineered the desired two-body dissipation in a scalable manner.

\begin{figure*}[t]
\centering
\includegraphics[width=17.2cm]{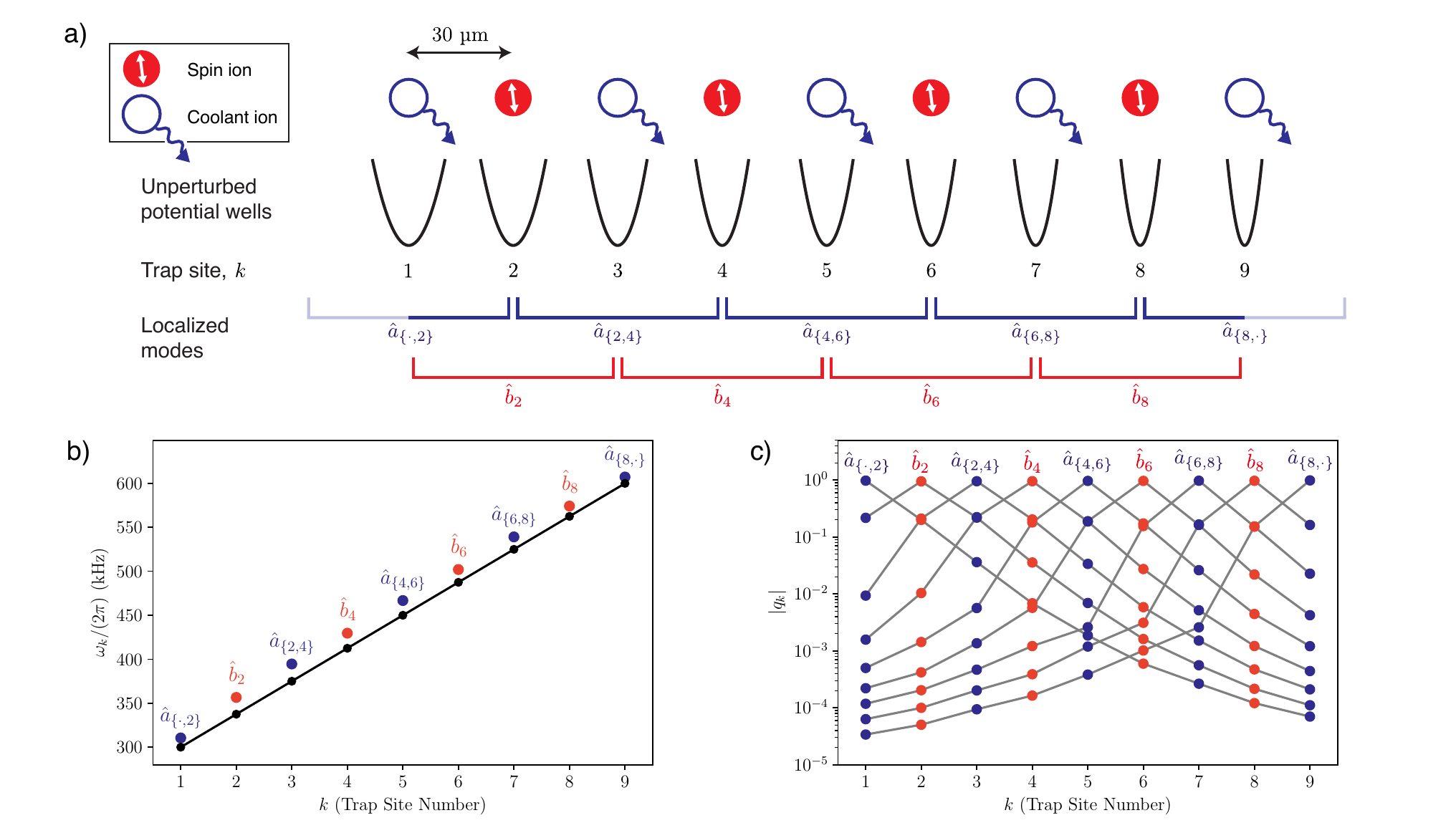}
\caption{
Mixed species implementation using micro-traps. (a). Ions in a linear chain of micro-traps are arranged along the axial direction (this corresponds to the axis with no radio frequency (r.f.)  curvature in the case of Paul traps, and equivalently, the magnetic field axis in the case of Penning traps). The distance between neighboring traps here is $30$~\um. Two kinds of ions are considered -- one to encode the spins (the spin species, here \cafthree), and the other to facilitate cooling (the coolant species, here \caf).  Each trap consists of a single ion with the species of the ion occupying the trap sites alternating between the spin and the coolant. By appropriately tuning the individual potential wells one can obtain motional modes localized to either one or two spin ions. (b). Frequency spectrum of the axial modes for an example configuration that leads to the localization of modes. In black are the axial frequencies of the individual micro-traps. These vary linearly with the trap site. The colored dots indicate the frequencies of the modes of collective motion for this system of ions. The motion in each mode is at a frequency close to the bare frequency of the trap and is dominated by the ion located at the trap (and its next-neighboring traps). (c). Eigenvectors of the axial modes showing the amplitudes of motion of each ion in the mode for this configuration. The participation of next-nearest (and further) neighboring ions drops off steeply in any given mode. Thus, as an approximation, each mode can be treated as being limited to triplets (or pairs in the case of ions at the edge) of ions.
}
\label{fig:microtrap}
\end{figure*}

Single-body dissipation in Eq.~\eqref{EqLindZ1}, is again realized -- now for the whole chain -- following the recipe in Sec. \ref{sec:repump}:
Using locally addressed repumper beams to an unstable level $\ket{r}$ for each individual ion, we achieve local jump operators
\begin{eqnarray}
L_j^{(1)}
= \sqrt{\gamma_1} \ket{\dn}_j \bra{\up} 
\equiv \sqrt{ \frac{\Gamma_{\dn r} \Omega_{\mathrm{rep},\up}^2}{\Gamma_r^2} } \ket{\dn}_j \bra{\up}.
\label{eq:Lrep:scale}
\end{eqnarray}
Associating $\gamma_1 = \epsilon(1-\gamma)$, we have thus realized the desired single-body dissipation (Eq.~\eqref{EqLindZ1}) for all ions in the chain.

The implementation of generalized dissipation in the $x$-basis, such as those in Eqs. \eqref{EqLindX1}--\eqref{EqLindX2} in Sec. \ref{sec:generalization} can be scaled up in an analogous manner.

\section{Normal mode engineering in an array of micro traps}
\label{sec:micro}

In the following, we discuss the physical implementation of the scalable setting detailed in Sec. \ref{sec:scalability} based on ion micro-traps \cite{Cirac2000, Wilson2014, Yang2016, Mielenz2016, Jain2018}. To implement the desired  mode structure, with localized modes subject to dissipation, we employ a novel approach to mixed-species normal mode engineering.

In an array of micro-traps individual control over the electric potential at each trap site allows us to engineer normal mode spectra suitable for implementing the desired operators. Previously the use of a special arrangement of the transverse frequencies of individual traps along a string has been proposed for encoding two-body bosonic gauge fields \cite{Yang2016}. Here we consider similar ideas to construct phonon modes localized to triplets of neighboring ions (that is any ion together with both of its nearest-neighbors), and with the use of mixed species of ions effectively encode two-body \emph{as well as} one-body operators. The use of two species of ions allows us to achieve this using mode engineering along only one axis of vibration instead of two transverse axes.

The calculation of the normal modes for a system of ions in an array of micro-traps \cite{98James1, Jain2018} involves first determining the equilibrium positions in the combined electric potential due to the trap electrodes and the Coulomb repulsion from the other ions. A Taylor series expansion about these positions up to second order then yields the effective harmonic potential experienced by each ion. The resulting Hamiltonian is then diagonalized to extract the eigenvalues, which give the frequencies of oscillation, and the eigenvectors, which give the relative amplitude of oscillation of each ion in any given normal mode. Since the trapping mechanism in Paul traps and Penning traps differs only in the radial directions, we focus here on the axial modes since the treatment would then be independent of the kind of trap for the discussion below.

To illustrate the idea we first consider a chain of a single species of trapped ions, each with mass $m$. The spatial separation between neighboring traps is given by $d$ and the potentials are designed so that the trap frequencies $\omega_k = \omega_0 + (k-1) \Delta_{\text{T}}$ increase linearly with the trap site $k$. If the frequency difference $\Delta_{\text{T}}$ is much larger than the two-ion exchange frequency, defined by $\Omega_{\text{ex}} = e^2 / \left(4 \pi \epsilon_0 m \omega_0 d^3 \right)$, we get for this system of coupled ions a spectrum of normal modes close to that of the non-interacting system. That is, the result is a set of modes with oscillation frequency near $\omega_k$ and participation mostly from the ion at that trap $k$ (as well as ions at the nearest-neighboring traps, $k-1$ and $k+1$).

Now consider a chain of alternating ionic species of nearly equal mass in these traps. One of these species serves as a coolant ion while the other one is used to encode the spins through two suitable internal states, and is from here on called the `spin ion'. The condition for nearly equal mass is useful for efficient sympathetic cooling of the spin species through the coolant species. As discussed above, the resulting normal mode structure consists of modes effectively localized to triplets of ions (except, of course, at the edges) but now there exist two kinds of modes -- one where the central ion is the coolant ion with two neighboring spin ions, and the other where the central ion is a spin ion between two coolant ions. Fig. \ref{fig:microtrap} a) shows this for a chain of alternating \caf and \cafthree ions arranged along the axial direction of the micro-trap array. The frequencies for the uncoupled arrangement of traps and for the coupled system are shown in figure \ref{fig:microtrap} b). Here $\omega_0 = 2 \pi \times 300$~kHz, $\Delta_{\text{T}} = 2 \pi \times 37.5$~kHz, and $d = 30$ ~\um. Since the participation of the ions drops roughly exponentially with the distance from the central trap we can assume almost no participation from other ions of the same species (since these lie two sites distant). This behavior can be seen from Fig. \ref{fig:microtrap} c), where the amplitudes of motion of each ion in the axial modes is plotted. Note that each mode is assigned a color so that it is clear which ions dominate the oscillation in a particular mode, and at what frequency. Modes $a$ which are localized to two spin-ions (and one coolant ion) allow us to engineer the two-body operators while modes $b$ which are localized to one spin-ion (and two coolant ions) allow for implementing one-body operators. To match the notation of the Sec. \ref{sec:scalability}, we associate the even-numbered trap sites with the spin ions by $2k \mapsto j$ (e.g. $a_{2,4}\mapsto a_{1,2}$), thereby excluding the odd-numbered trap sites containing the coolant ions. Employing sympathetic cooling of these modes, dissipation can be engineered in the $\sigma_x$-eigenbasis, following the recipes in Sec. \ref{sec:generalization}.

Note that the laser-ion couplings in Sec. \ref{sec:scalability}, Eqs. \eqref{eq:Hdrive:scale1}--\eqref{eq:Hsideband:scale3} require addressing of pairs of ions coherently with the same detuning $\Delta$. As discussed in Sec. \ref{sec:mechanism}, $\Delta$ needs, however, to be matched to the phonon detuning $\delta$, which differs from ion to ion because to the trap frequency offset. The resulting mismatch can be compensated by local AC Stark shifts on $\ket{e}$ and $\ket{f}$, which can be generated by individually addressed lasers.

\subsection{Alternative realization in linear ion traps}
\label{sec:platforms}

In conventional bulk ion traps, the desired dynamics can be implemented in a stepwise manner. Here we consider a sequential realization based on delocalized motional modes in combination with local addressing techniques \cite{Schindler2013, Jurcevic2014, Warring2013, Debnath2016, Landsman2019}.
We assume all ions to be detuned by a constant amount with respect to the coupling configurations in Sec. \ref{sec:scalability}. For $N$ ions, we now consider $N$ timesteps. During each step, we direct a pair of individual addressing lasers on a pair of ions. This beam shifts the transitions of the ions into resonance with the carrier and sideband couplings in Eqs. \eqref{eq:Hdrive:scale1}--\eqref{eq:Hsideband:scale3}. We thereby pairwise realize the desired dynamics on ions $j$ and $j+1$, leaving the other ions uncoupled. In the next step, the individual addressing laser is shone onto another pair of ions, $j+1$ and $j+2$, and so forth. Provided short modulation times for the lasers, the timeframes for the ions may be reduced to stroboscopic length, resulting in a ``Trotterized'' realization of the desired dynamics.

\section{Conclusion and Outlook}

We have presented a scheme suitable to revive the effects of integrability in a controllably driven and open setup, where the underlying Hamiltonian dynamics is only approximately integrable.
The scheme is based on weak couplings to Markovian baths in combination with nearly integrable quantum spin Hamiltonians, ingredients which are readily available in state-of-the-art trapped-ion setups. In addition, we present a novel technique for the engineering of motional modes in an array of mixed-species micro-traps, which support the realization of the desired dynamics.

Our numerical analysis shows that despite different sources of integrability breaking due to long-range interactions in the Hamiltonian and openness itself, a steady state is realized that cannot be modeled as a thermal ensemble. Instead, approximate expectation values of local observables can be obtained from a generalized Gibbs ensemble and we identify the experimental signatures which reveal that. 
We presented results for a rotated XY and transverse-field Ising Hamiltonian; however, the same Lindblad operators would activate a GGE in the interacting XXZ Heisenberg chain.  

While our goal has been to engineer Lindblad dissipators which stabilize a non-trivial and highly non-thermal steady state by weak driving, a simpler task would be to consider a non-trivial dynamics towards a trivial (thermal, maximally mixed or empty) steady state. As has been shown in Ref.~\cite{lange18}, the dynamics of a (nearly) integrable system weakly coupled to {\it arbitrary} Lindblad baths can be approximately described with a {\it time-dependent GGE}. An example of this is atom loss in cold-atom setups, whose effects have been observed \cite{johnson17} and theoretically addressed \cite{bouchoule20} very recently. The simplest combination of the strategy presented above would be to follow the dynamics of an XY-Hamiltonian, Eq.~\eqref{EqXYh}, in the presence of single-body decay, Eq.~\eqref{EqLindZ1}, alone.

Beyond this work, our dissipation engineering strategies open the door to experiments that will shed light on novel phenomena in open quantum systems. 
For example, the dissipators presented here have been used to study many-body localization through the perspective of open systems \cite{lenarcic19}.
Our novel mixed-species mode engineering techniques based on arrays of ion micro-traps hold promise to become a powerful tool in quantum simulation. Here, sympathetic cooling is not only useful to reduce the entropy of the system, but also to realize complex dissipators. Generalizing these techniques may allow addressing open questions in non-equilibrium quantum many-body physics.

\section{Acknowledgments}
We thank Achim Rosch, Petar Jurcevic, Christine Maier, and Christian Roos for valuable discussions. We acknowledge financial support from the Swiss National Science Foundation through the National Centre of Competence in Research for Quantum Science and Technology (QSIT) grant 51NF40-160591. FR acknowledges financial support from the Alexander von Humboldt foundation through a Feodor Lynen fellowship and from the Swiss National Science Foundation (Ambizione grant no. PZ00P2$\_$186040). FL acknowledges the financial support of the German Science Foundation under CRC TR 183 (project A01). ZL was financially supported by Gordon and Betty Moore Foundation’s EPIC initiative, Grant No. GBMF4545 and from the European Research Council (ERC) synergy UQUAM project, and partially by TR CRC 183 (project A01). ZL also acknowledges the L'Or\'eal-Unesco national scholarship 
``For women in science" and the NeTex program of University of Cologne which funded visits of Harvard University where this work was initiated.

\appendix
\section{Theory of weakly driven nearly integrable systems}
\label{sec::Ensembles}
Here we review the theory of weakly driven and open nearly integrable systems. We consider the setup discussed in the main text, where the dynamics of the system is given by a dominant integrable Hamiltonian $H_0$, in the presence of perturbations of unitary and Markovian nature, which weakly break the integrability. The corresponding Liouvillian terms are 
\begin{align}
\cl{L}_0 \rho &= -i[H_0,\rho], \notag  \\ 
\cl{L}_u \rho &= -i\left[H_1(\epsilon_1),\rho\right],\notag  \\
\cl{\mathcal{L}}_m\rho&=  \left( \cl{D}^{(1)}(\epsilon) +  \cl{D}^{(2)}(\epsilon) \right) \rho.\notag 
\end{align}
The steady state density matrix is determined by
\begin{equation}\label{eq::SS}
\dot{\rho}_\infty=(\cl{L}_0 + \cl{L}_p)\rho_\infty=0,
\end{equation}
where $\cl{L}_p=\cl{L}_u + \cl{L}_{m}$.
Because perturbations due to the Markovian dissipation $\cl{L}_m$ and the next-nearest neighbor interaction $\cl{L}_u$ are only weak, the exact steady state density matrix $\rho_{\infty}$ can be split as
\begin{equation}\label{EqRhoNESS}
\rho_{\infty}= \rho_{\textrm{BD}} + \delta\rho,
\end{equation}
where a $\rho_{\textrm{BD}}\sim \mathcal{O}(1)$, while $\delta\rho(\epsilon, \epsilon_1)$ is a small correction regulated by the strength of unitary and Markovian perturbations.
$\rho_{\textrm{BD}}$ must fulfill the zero-th order Liouville stationarity equation,
\begin{equation}
\cl{L}\rho_\infty \approx \cl{L}_0 \rho_{\textrm{BD}}=-i[H_0,\rho_{\textrm{BD}}]=0	
\end{equation}
and therefore must have a block-diagonal form
\begin{equation}\label{EqRho0}
 \rho_{\textrm{BD}}=\sum_{m,n} a_{mn} |m\rangle \langle n| \, \delta_{E^0_m,E^0_n}, \quad 
 H_0\ket{n}=E_n^0 \ket{n}
 \end{equation} 
with respect to the eigenstates of $H_0$ and is parametrized with about $2^N$ parameters $a_{mn}$.

However, it turns out that description with $\rho_{\textrm{BD}}$ is redundant and can be replaced with a generalized Gibbs ensemble, 
$\rho_{\textrm{BD}} \rightarrow \rho_{\textrm{GGE}}$,
if $H_0$ is integrable \cite{lange17,lange18} and with a Gibbs ensemble 
$\rho_{\textrm{BD}} \rightarrow \rho_{\textrm{th}}$,
if $H_0$ is ergodic \cite{shirai18}.

While equivalence of $\rho_{\textrm{BD}}$ and $\rho_{\textrm{GGE}}$ is formally expected when calculating expectation values of local observables in the thermodynamic limit and with all conservation laws included, 
in most cases also a truncated GGE (tGGE) with a few conservation laws
\begin{equation}\label{EqRhotGGE}
\rho_{\textrm{tGGE}} \equiv \frac{e^{-\sum_{i=1}^{N_C} \lambda_i C_i}}{\tr[e^{-\sum_{i=1}^{N_C} \lambda_i C_i}]}
\end{equation}
qualitatively well captures the expectation values of local observables, if their support is much smaller than the support of included conservation laws. Known exceptions are observables that are orthogonal to all included conservation laws \cite{prosen11}.

In the following we will compare the expectation values evaluated with respect to 
$\rho_\infty$, $\rho_{\textrm{BD}}$ and $\rho_{\textrm{tGGE}}$ in order to confirm that steady states can be approximately described with non-thermal generalized Gibbs ensembles.
Note that since our choice of Lindblad operators breaks magnetization, as well as momentum conservation, the whole Hilbert space is of relevance.

The Lagrange parameters $\lambda_i$ in Eq.~\eqref{EqRhotGGE} and parameters $a_{mn}$ in Eq.~\eqref{EqRho0} are determined from the stationarity conditions in the steady state \cite{lange17,lenarcic18,lange18}, 
\begin{align}
\partial_t \tr[C_i\, \rho_{\textrm{tGGE}}]&\overset{!}{=}0, \\
\partial_t \tr[\ket{m}\bra{n}\rho_{\textrm{BD}}]&\overset{!}{=}0, \quad E^0_m=E^0_n
\end{align}
for $\rho_{\textrm{tGGE}}$ and $\rho_{\textrm{BD}}$, respectively. 
For our choice of perturbation the contributions to order $\epsilon$ and $\epsilon_1^2$
\begin{align}\label{EqCondA}
\ave{\dot{C}_i} & \approx \tr[(\cl{L}_m + \cl{L}_{u,2})\rho_{\textrm{tGGE}}] \overset{!}{=} 0,\\
\cl{L}_{u,2} &\equiv - \cl{L}_u \cl{L}_0^{-1} \cl{L}_u
\end{align} 
uniquely fix the $\lambda_i$ (or equivalently $a_{mn}$) in the steady state.
Note that $\cl{L}_0 \rho_{\textrm{tGGE}}=0$ because $[H_0,C_i]=0$, and that unitary perturbation contributes to the decay of conservation laws only in the second order, since $\tr[C_i \cl{L}_u \rho_{\textrm{tGGE}}]=0$ due to cyclicity of the trace. 
More details on the derivation of condition \eqref{EqCondA} and how to use $\cl{L}_0^{-1}$ in practice can be found in Ref. \cite{lenarcic18}. Here we give only the final result for $\ave{\dot{C}_i}$,
\begin{align}
\ave{\dot{C}_i}=2 & \pi \sum_{nm} 
(\ave{m|C_i|m}-\ave{n|C_i|n}) \ave{n|\rho_{\textrm{tGGE}}|n} \label{eq::dCi}\\
\times \Big[
&|\ave{n|H_1|m}|^2 \, \frac{1}{\pi}\frac{\eta}{(E_n^0 - E_m^0)^2 + \eta^2} \notag\\
+\sum_{j} &\left((1-\gamma) \left|\ave{n|L_j^{(1)}|m}\right|^2+ \gamma \left|\ave{n|L_j^{(2)}|m}\right|^2\right)
\Big] \notag
\end{align}
where finite broadening $\eta$ has to be used for calculations at finite system sizes. Expressions relevant for $\rho_{\textrm{BD}}$ are obtained by replacements $C_i \to \ket{m}\bra{n}$ and $\rho_{\textrm{tGGE}}\to \rho_{\textrm{BD}}$.

\subsection{Numerical results}\label{sec::results}
We base our analysis on three approaches: (i) calculation of the exact steady state $\rho_\infty$, Eq.~\eqref{eq::SS}, at finite but small $\epsilon$, obtained from diagonalization of the full Liouvillian on small system sizes $N=6$ where we exclude or include (NN) unitary integrability breaking $H_1$, (ii) exact calculation of $\rho_{\textrm{BD}}$, Eq.~\eqref{EqRho0}, 
on $N=6,8,10$ and (iii) approximate calculation based on a truncated GGE, Eq.~\eqref{EqRhotGGE}, including a finite number of $N_C=4$ conservation laws $C_i$ on $N=10$.
Note that each $C_i=\sum_i c^{(i)}_j$ is a translationally invariant sum of operators $c^{(i)}_j$ with support not larger than $i$. Due to finite size effects, only $C_i$ with support smaller than $ N/2$ can be included in the tGGE.
$C_i$ are obtained using the so-called boost operator, $B=-i\sum_j j h_j$, where $H_0= \sum_j h_j$, from the recursive relation $C_{i+1} = [B,C_i]$ for $i \geq 2$ and $C_2=H_0$. At the isotropic point, $J_y=J_z$ the magnetization $S^x=C_1$ is conserved as well.

\begin{figure}[b!]
\includegraphics[width=.9\linewidth]{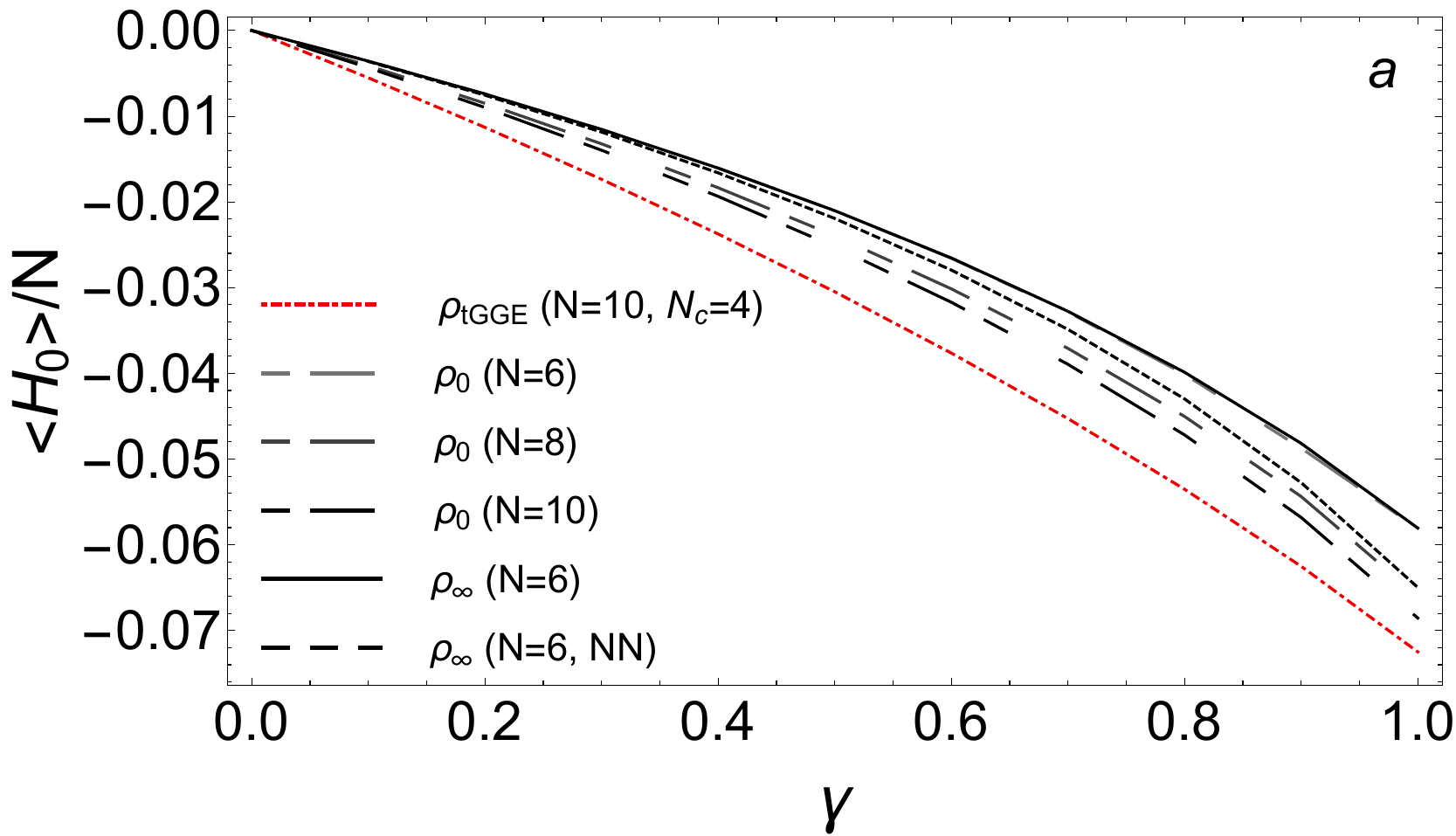}
\includegraphics[width=.9\linewidth]{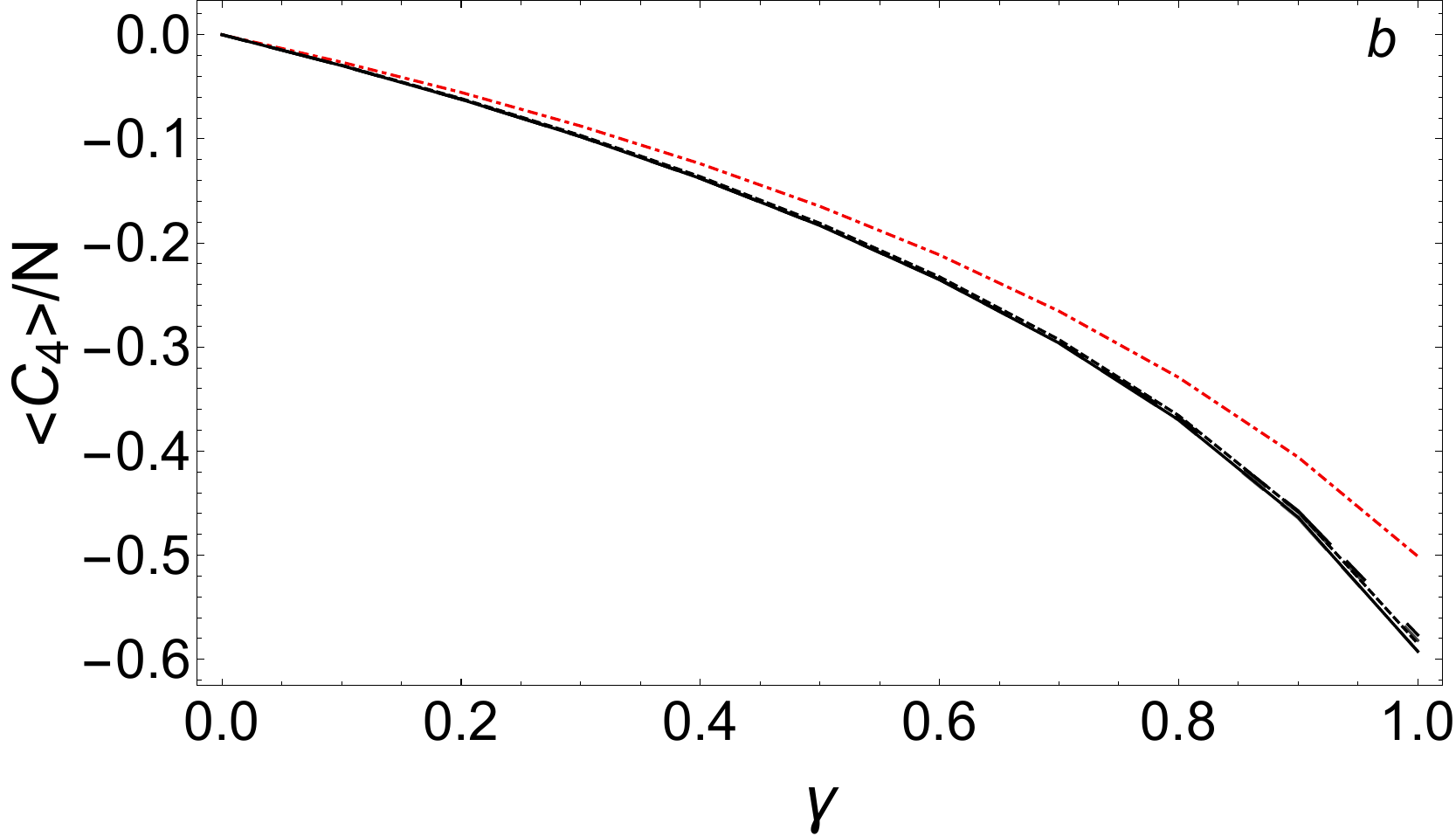}
\caption{Expectation values of (a) energy and (b) $C_4$ densities as a function of relative driving strength $\gamma$, Eqs.~(\ref{EqLindZ1},\ref{EqLindZ2}). We compare results calculated i) with (NN) or without next nearest coupling $H_1$ from the exact steady state $\rho_{\infty}$ at $\epsilon=0.01$, ii) from ansatz $\rho_{\textrm{BD}}$, and iii) from a truncated GGE $\rho_{\textrm{tGGE}}$, using $N_C=4$ conservation laws. System sizes $N=6,8,10$ are used at $J_y=h=1$, $J_z=0.1$ and $\epsilon_1=0.05$ for (NN).
\label{FigExpValApp}}.
\end{figure}

Fig.~\ref{FigExpValApp} shows $\ave{H_0}$ and $\ave{C_4}$ as a function of relative dissipator strength $\gamma$, Eqs.~(\ref{EqLindZ1},\ref{EqLindZ2}), obtained using different approximations described above at largest accessible system sizes. We observe a good agreement between the three approaches, also for other parameters not displayed.
Results calculated from $\rho_{\textrm{BD}}$ on $N=6,8,10$ interpolate between the exact ($N=6$) and tGGE ($N=10$) results. While $\rho_{\textrm{BD}}$ and $\rho_{\infty}$ for small $\epsilon=0.01$ agree very well on $N=6$, increasing the system size shows a tendency of $\rho_{\textrm{BD}}$ towards the $\rho_{\textrm{tGGE}}$ result. A milder discrepancy of $\rho_{\textrm{tGGE}}$ results is due to omitted conservation laws. 

The important conclusion is two-fold: (i) The above analysis gives numerical support for the claim that the stabilized steady state can be approximated with a GGE despite different sources of integrability breaking. (ii) While $\rho_{\textrm{tGGE}}$ is parametrized with $N_C=4$ parameters, $\rho_{\textrm{BD}}$ at $N=10$ with about $10^3$ and the full $\rho_\infty$ would require about $10^6$ parameters. Description in terms of a truncated GGE is therefore a highly compact parametrization of the steady state, which takes into account only the most relevant information.

We find that in the presence of Lindblad driving, the effect of next-nearest interaction, $H_1$, is rather weak. While in a closed setup $H_1$ is crucial as it dictates relaxation towards a thermal state, it is dominated with Lindblad terms in an open setup. Mathematically, this can be explained through Eq.~\eqref{eq::dCi} which shows that the unitary perturbation is constrained to act only between degenerate eigenstates, while the Markovian contribution has no such constraint.
Fig.~\ref{FigExpValApp} shows that results obtained from the exact steady state $\rho_{\infty}$ calculated with (NN) or without $H_1$ are very similar. While $H_1$ can be easily included into the calculation of the exact steady state $\rho_\infty$, it brings certain ambiguity into the calculation of $\rho_{\textrm{BD}}$ and $\rho_{\textrm{tGGE}}$. Namely, on finite system sizes one has to introduce broadening $\eta$ when calculating $\cl{L}_u \cl{L}_0^{-1}\cl{L}_u$, Eq.~\eqref{eq::dCi}. As we showed in \cite{lange18}, broadening itself modifies the effective strength of the perturbation, meaning that different system sizes, requiring different broadening, cannot be directly compared. Since $\rho_\infty$ shows that the effect of $H_1$ is small, we omit it in the calculation of $\rho_{\textrm{BD}}$ and $\rho_{\textrm{tGGE}}$.

\section{Microscopic derivation of two-body decay}
\label{sec:analysis}

In the following, we verify that the mechanisms presented in Sec. \ref{sec:mechanism} lead to the desired dissipative couplings in Eq.~\eqref{EqLindZ2}. To this end, we eliminate the excited degrees of freedom by means of the \textit{effective operator formalism} \cite{Reiter12}. This allows us to obtain the effective dynamics of the ground states.

To obtain the effective processes between the ground states, we need to evaluate the expressions for the effective Hamiltonian and Lindblad operators \cite{Reiter12},
\begin{align}
\label{eq:Heff}
H_{\mathrm{eff}} &= - \frac{1}{2} \left( V_- H_\mathrm{NH}^{-1} V_+ + H.c. \right),
\\
L_{\mathrm{eff},k} &= L_k H_\mathrm{NH}^{-1} V_+,
\label{eq:Leff}
\end{align}
with the relevant terms discussed below.

For the scheme at hand, $V_+$ is the weak excitation from the ground states to the excited states (de-excitation: $V_- = V_+^\dagger$), taken from Eq.~\eqref{eq:Hdrive},
\begin{align}
V_+ = \frac{\Omega}{2} ( \sqrt{2} \ket{\psi_e} \bra{\dn\dn} + \ket{\up e} \bra{\up \dn} ).
\end{align}
While $L_k$ can represent various sources of dissipation, the only relevant jump operator is given by Eq.~\eqref{eq:Lrep:e}, which can be written as
\begin{align}
L_{\Gamma} = \sqrt{\frac{\Gamma}{2}} \ket{\up\dn} \bra{\psi_e}.
\label{eq:Lexc}
\end{align}
The evolution of the excited states is described by a non-Hermitian Hamiltonian,
\begin{align}
\label{eq:HNH}
H_\mathrm{NH} = H_e - \frac{i}{2} \sum_k L_k^\dagger L_k= H_e - \frac{i}{2} L_\Gamma^\dagger L_\Gamma,
\end{align}
incorporating the excited-state Hamiltonian $H_\mathrm{e} = H_\mathrm{e,\dn\dn} + H_\mathrm{e,\up\dn}$, with $H_\mathrm{e,\dn\dn}$ as of Eq.~\eqref{eq:He00} and
\begin{align}
\label{eq:He10}
H_\mathrm{e,\up\dn}
&= \Delta \ket{\up e} \ket{0} \bra{0} \bra{\up e} + \delta \ket{\up\dn} \ket{1} \bra{1} \bra{\up\dn}
\\ \nonumber
&+ g ( \ket{\up e} \ket{0} \bra{1} \bra{\up\dn} + \ket{\up\dn} \ket{1} \bra{0} \bra{\up e}).
\end{align}
The jump operators relevant for Eq.~\eqref{eq:HNH} are given by induced spontaneous emission, as described by Eq.~\eqref{eq:Lexc}.
The non-Hermitian terms in Eq.~\eqref{eq:HNH} can then be taken into account by generalizing the detunings from $H_e$ to ``complex'' energies of the form $\tilde{\Delta} = \Delta - i (\Gamma / 2) / 2$. Here we assume no motional decoherence and hence, $\tilde{\delta} = \delta$. If necessary, processes like phonon decay, $L_\kappa = \sqrt{\kappa} a$, can be taken into by $\tilde{\delta} = \delta - i \kappa / 2$. We obtain $H_\mathrm{NH} = H_\mathrm{NH, \dn\dn} + H_\mathrm{NH, \up\dn}$, with
\begin{align}
\label{eq:HNH00}
H_\mathrm{NH,\dn\dn}
&= \tilde{\Delta}_{\dn\dn} \ket{\psi_e} \ket{0} \bra{0} \bra{\psi_e} + \tilde{\delta} \ket{\dn\dn} \ket{1} \bra{1} \bra{\dn\dn}
\\ \nonumber
&+ \sqrt{2} g ( \ket{\psi_e} \ket{0} \bra{1} \bra{\dn\dn} + \ket{\dn\dn} \ket{1} \bra{0} \bra{\psi_e}).
\\
\label{eq:HNH10}
H_\mathrm{NH,\up\dn}
&= \tilde{\Delta}_{\up\dn} \ket{\up e} \ket{0} \bra{0} \bra{\up e} + \tilde{\delta} \ket{\up\dn} \ket{1} \bra{1} \bra{\up\dn}
\\ \nonumber
&+ g ( \ket{\up e} \ket{0} \bra{1} \bra{\up\dn} + \ket{\up\dn} \ket{1} \bra{0} \bra{\up e}),
\end{align}
having defined $\tilde{\Delta}_{\dn\dn} = \Delta - i (\Gamma / 2) / 2$, $\tilde{\Delta}_{\up\dn} = \Delta$, $g_{\dn\dn} = g$, and $g_{\up\dn} = \sqrt{2} g$.
$H_\mathrm{NH}$ is block-diagonal and hence simple to invert,
\begin{align}
H_\mathrm{NH}^{-1} &= H_{\mathrm{NH},\dn\dn}^{-1} + H_{\mathrm{NH},\up\dn}^{-1},
\\
H_{\mathrm{NH},\dn\dn}^{-1} &= \tilde{\Delta}_{\dn\dn,\mathrm{eff}}^{-1} \ket{\psi_e} \ket{0} \bra{0} \bra{\psi_e} + \tilde{\delta}_{\dn\dn,\mathrm{eff}}^{-1} \ket{\dn\dn} \ket{1} \bra{1} \bra{\dn\dn}
\nonumber \\ \nonumber
&+ g_{\dn\dn,\mathrm{eff}}^{-1} ( \ket{\psi_e} \ket{0} \bra{1} \bra{\dn\dn} + \ket{\dn\dn} \ket{1} \bra{0} \bra{\psi_e}), 
\\
H_{\mathrm{NH},\up\dn}^{-1} &= \tilde{\Delta}_{\up\dn,\mathrm{eff}}^{-1} \ket{\up e} \ket{0} \bra{0} \bra{\up e} + \tilde{\delta}_{\up\dn,\mathrm{eff}}^{-1} \ket{\up\dn} \ket{1} \bra{1} \bra{\up\dn}
\nonumber \\ \nonumber
&+ g_{\up\dn,\mathrm{eff}}^{-1} ( \ket{\up e} \ket{0} \bra{1} \bra{\up e} + \ket{\up\dn} \ket{1} \bra{0} \bra{\up e}).
\end{align}
Here we have defined effective detunings and couplings,
\begin{align}
\tilde{\Delta}_{ij,\mathrm{eff}} &= \tilde{\Delta}_{ij} - \frac{g_{ij}^2}{\tilde{\delta}},
\\
\tilde{\delta}_{ij,\mathrm{eff}} &= \tilde{\delta} - \frac{g_{ij}^2}{\tilde{\Delta}_{ij}},
\\
\tilde{g}_{ij,\mathrm{eff}} &= g_{ij} - \frac{ \tilde{\Delta}_{ij} \tilde{\delta} }{ g_{ij} },
\end{align}
which mediate the effective processes.
Using Eq.~\eqref{eq:Leff}, we obtain for the effective jump operators for spontaneous emission
\begin{align}
\label{eq:LeffGammaApp}
L_{\mathrm{eff}}^{(2)} \equiv L_{\mathrm{eff},\Gamma} = \sqrt{\gamma_2} \ket{\up\dn}\ket{0} \bra{0}\bra{\dn\dn},
\end{align}
with the effective decay rate
\begin{align}
\gamma_2 &= \frac{ (\Gamma/2) (\Omega/\sqrt{2})^2}{|\tilde{\Delta}_{\dn\dn,\mathrm{eff}}|^2}.
\end{align}
For the parameter choice of Sec. \ref{sec:mechanism} ($\Delta = \delta = \sqrt{2} g$), we find
\begin{align}
\tilde{\Delta}_{\dn\dn,\mathrm{eff}} &= \tilde{\Delta}_{\dn\dn} - \frac{2 g^2}{\tilde{\delta}} = - \frac{ i \Gamma }{4},
\end{align}
This yields an effective decay rate
\begin{align}
\gamma_2 &\approx \frac{4 \Omega^2}{\Gamma}.
\label{eq:gammaplus1}
\end{align}
We can now associate the effective Lindblad operator in Eq.~\eqref{eq:LeffGammaApp} with the desired one in Eq.~\eqref{EqLindZ2},

\section{AC Stark shift of states}
\label{sec:ACStark}
We also derive the effective Hamiltonian using Eq.~\eqref{eq:Heff}, and obtain
\begin{align}
H_\mathrm{eff} 
&= - \frac{ ( \Omega / 2 )^2 }{ | \tilde{\Delta}_\mathrm{\up\dn, eff} |^2 } \mathrm{Re}(\tilde{\Delta}_\mathrm{\up\dn, eff}) \ket{\up\dn}\ket{0} \bra{0} \bra{\up\dn},
\label{eq:Heff2}
\end{align}
where $\mathrm{Re}()$ denotes the real part. For our parameter choice $\Delta = \delta = \sqrt{2} g$, we have
\begin{align}
\tilde{\Delta}_{\up\dn,\mathrm{eff}} &= \tilde{\Delta}_{\up\dn} - \frac{g^2}{\tilde{\delta}} = \frac{g}{\sqrt{2}},
\end{align}
and, thus,
\begin{align}
H_\mathrm{eff} 
&= - \frac{ \Omega^2 }{ 2 \sqrt{2} g } \ket{\up\dn}\ket{0} \bra{0} \bra{\up\dn},
\label{eq:Heff3}
\end{align}
This term corresponds to an AC Stark shift of $\ket{\up\dn}\ket{0}$, which can be safely neglected in the considered parameter regime $\Omega^2 \ll \{\Gamma^2, g^2\}$.

\bibliography{./references}

\end{document}